\documentclass[eps,12pt]{report}
\usepackage{graphicx}
\RequirePackage{amssymb}
\RequirePackage{setspace}   
\RequirePackage{graphicx}   
\RequirePackage{lscape}     
\RequirePackage{amssymb}    
\RequirePackage{longtable}  
\RequirePackage{fancyhdr}   
\RequirePackage{xspace}     
\RequirePackage{multicol}   
\RequirePackage{fancybox}   
\RequirePackage{calc}

\DeclareRobustCommand{\ion}[2]{\relax\ifmmode
                              \ifx\testbx\f@series
                              {\mathbf{#1\,\mathsc{#2}}}\else
                              {\mathrm{#1\,\mathsc{#2}}}\fi
                              \else\textup{#1\,{\mdseries\textsc{#2}}}%
                              \fi}

\newcommand{\thetitle}{Rotation and color properties of the nucleus of comet 2P/Encke}
\begin{document}
%\begin{sloppypar}
% TITLE PAGE (1)%%%%%%%%%%%%%%%%%%
\begin{center}
\textbf{\thetitle}\\
\vspace{2.5cm}
\begin{small}
Stephen C. Lowry\\
Queen's University Belfast, 
Astrophysics Research Center,\\
School of Mathematics and Physics,
Belfast, BT7 1NN, United Kingdom.\\
Email: s.c.lowry$@$qub.ac.uk\\
Paul R. Weissman\\
Jet Propulsion Laboratory, MS 183-301, 4800 Oak Grove Drive, Pasadena, CA 91109, USA.\\
\end{small}
\end{center}

%\vspace{1.0cm}
%\begin{center}
%\begin{large}
%\textbf{Icarus Manuscript No. I09753: Accepted version}
%\end{large}
%\end{center}

\vspace{1.0cm}
\begin{center}
\begin{large}
\textbf{Accepted for publication in Icarus}
\end{large}
\end{center}

\vspace{3.0cm}
\noindent
Pages:   27\\
Tables:  03\\
Figures: 07\\

%%%%%%%%%%%%%%%%%%%%%%%%%%
\newpage
\noindent
\textbf{Proposed Running Head:} Rotation and color properties of comet 2P/Encke \\ 

\vspace{2.5cm}
\noindent
\textbf{Editorial correspondence to:}\\
Dr. Stephen C. Lowry\\
Queen's University Belfast\\
Astrophysics Research Centre\\ 
School of Mathematics and Physics\\
Belfast, BT7 1NN, United Kingdom.\\
Phone: 44 28 97273692\\
E-mail: s.c.lowry@qub.ac.uk\\
\clearpage
%%%%%%%%%%%%%%%%%%%%%%%%%%%%%%%%%%%%%%%%%%%%5

\noindent
\begin{center}
\textbf{Abstract}
\end{center}

We present results from CCD observations of comet 2P/Encke 
acquired at Steward Observatory's 2.3m Bok Telescope on Kitt Peak. 
The observations were carried out in October 2002 when the comet 
was near aphelion.
Rotational lightcurves in $B$, $V$ and $R$-filters were acquired over two 
nights of observations, and analysed to study the physical and color properties
of the nucleus.
The average apparent $R$-filter magnitude across both nights 
corresponds to a mean effective radius of
$3.95 \pm 0.06$ km, and this value is similar to that found for the $V$ and $B$ filters.
Taking the observed brightness range, we obtain 
$a/b \geq 1.44 \pm 0.06$ for the semi-axial ratio of Encke's nucleus. 
Applying the axial ratio to the $R$-filter photometry gives nucleus semi-axes of 
[$3.60 \pm 0.09$]$\times$[$5.20 \pm 0.13$] km, using the empirically-derived albedo
and phase coefficient. 
No coma or tail was seen despite deep imaging of the comet,
and flux limits from potential unresolved coma do not exceed a few percent of
the total measured flux, for standard coma models. 
This is consistent with many other published data sets taken when the comet 
was near aphelion. 
Our data includes the first detailed time series multi-color measurements of a 
cometary nucleus, and significant color variations were seen on October 3, though
not repeated on October 4.
The average color indices across both nights are:
($V-R$) = $0.39 \pm 0.06$ and ($B-V$) = $0.73 \pm 0.06$ ($\bar{R}$ = $19.76 \pm 0.03$).
We analysed the $R$-filter time-series photometry using the method
of Harris et al. (1989) to constrain the rotation period of the comet's nucleus,
and find that a period of $\sim$11.45 hours will satisfy the data, however the errors bars
are large. 
We have successfully linked our data with the September 2002 data from Fern\'{a}ndez et al. (2005) 
- taken just 2-3 weeks before the current data set - and we show that a rotation period of 
just over 11 hours does indeed work extrememly well for the combined data set. The resulting 
best fit period is $11.083 \pm 0.003$, consistent with the Fern\'{a}ndez et al. value.

\vspace{1.5cm}

%%%%%%%%%%%%%%%%%%%%%%%%%%%%%%%%%%%%%%%%%%%%5
%\newpage
\noindent
\section{\bf Introduction}
\label{INTRODUCTION}

Probing the physical properties of cometary nuclei is a challenge due
to the presence of dust and gas comae that envelope the nuclei at small
heliocentric distances. 
Cometary nuclei represent the least altered material from which to constrain 
the conditions and formation mechanisms of the early solar system as well as various 
subsequent evolutionary processes.
By far the most well studied cometary population are the Jupiter-family comets (JFCs). This 
population are the most accessible due to their proximity to Earth and because
many have small or negligible surface outgassing beyond 3 AU from the Sun,
allowing the reflected and/or thermally emitted flux from the nucleus to be detected directly. 
The formation location of JFCs is most likely the trans-Neptunian region 
(Duncan et al. 1995; Ip and Fern\'{a}ndez 1997).
In contrast, the nuclei of long-period comets and Halley-type comets most likely formed  
closer to the Sun in the region of the giant planets. 

Studying the very low activity, near-dormant JFC comets provides a means for understanding
the end states of comets. Indeed, comet Encke - a highly evolved member of the JFCs 
and the focus of this paper -  may be close to becoming an inert comet. 
Thus it is important to study the 
surface and bulk physical properties of its nucleus. A systematic survey of such 
transitional objects will allow comparisons to be made with the so-called
ACO population (asteroids in cometary orbits), or that of the near-Earth asteroids,
to establish possible physical links.

Full physical characterization, including determination of spin-rates has been accomplished 
for only a few comets. For a recent and thorough review of this topic see Lamy et al.
(2004) as well as reviews by Samarasinha et al. (2004) and Weissman et al. (2004). 
Rotation statistics of comets are important for studying the physical
evolution of these bodies and how they compare with their suspected
progenitor population, the Kuiper belt objects. These distributions are expected to 
be different, especially due to outgassing torques that act on the nuclei as they
enter the inner Solar system. Nucleus rotation and shape information can be obtained 
from time-series imaging of the comet, but unfortunately full nucleus 
lightcurves have been obtained for very few comets, due to the difficulties noted above. 
With full rotation lightcurves one also obtains a better measurement of the mean size 
of the nucleus, which aids in deriving a more accurate size distribution.

Simultaneous multi-color lightcurves have not been obtained for any cometary
nuclei and so the only data we have that points towards surface inhomogeniety 
comes from in-situ data acquired from spacecraft flybys, including Giotto, Vega, 
Deep Space 1, Stardust, and Deep Impact. When 19P/Borrelly was
imaged by the Deep Space 1 probe, the surface normal-reflectance was seen to vary  
from 0.01--0.05 (Buratti et al. 2004).

Comet Encke has the shortest orbital period of any known comet (3.3 years), 
and its aphelion distance of 4.1 AU is well within the orbit of Jupiter. 
This orbital stability and short periodicity has allowed observations
over many apparitions. Encke is therefore one of the most well studied comets, 
and is one of the few comets for which the albedo and phase coefficient have
been measured (Fern\'{a}ndez et al. 2000). The nucleus has one of the steepest
phase darkening slopes measured for a cometary nucleus with a phase coefficient
of 0.06 magnitudes/degree, measured over the phase angle range of $0-106^{\circ}$. 
The visual geometric albedo is also relatively high at 
$0.05 \pm 0.02$, although the uncertainty overlaps the commonly accepted
canonical albedo of 0.04. 

Earlier attempts were made at constraining Encke's rotation period.
Time-series optical photometry from Jewitt and Meech (1987)
show a most likely period of $22.43 \pm 0.08$ hours, whereas
Luu and Jewitt (1990) found a best-fit period of $15.08 \pm 0.08$ hours. 
Both papers indicate that other periods were also consistent with the data.
Thermal infrared time-series observations by Fern\'{a}ndez et al. (2000) 
are also consistent with the $15.08$ hour period.
A large data set was presented by Fern\'{a}ndez et al. (2005), acquired 
from July 2001 to September 2002, during which time the comet was near aphelion.
They found that the synodic spin-rate is either $11.079 \pm 0.009$ hours or
$22.158 \pm 0.012$, and that these periods were incompatible with the earlier 
estimates from Jewitt and Meech (1987) and Luu and Jewitt (1990), and vice versa.
Belton et al. (2005) performed a detailed analysis of the available $R$-band and 
10$\mu$m photometry and suggested that the nucleus may be in a complex or excited
rotation state. Out of the two possible short-axis mode (SAM) states that provide good fits to the
periodicities, Belton et al. found that the most likely state has a precessional frequency 
for the long axis about the total angular momentum vector $P_{\phi}$ of 11.1 hours, 
and an oscillation frequency around the long axis $P_{\psi}$ of 47.8 hours.

The Encke nucleus was observed using the Arecibo radar during the close approach in November
2003 and is thus the first comet to yield radar detections at multiple apparitions 
(Kamoun et al. 1982; Harmon and Nolan 2005). The new radar data supports the recently reported
11 hour rotation period, and excludes the longer 15 and 22 hour periods that 
were previously suggested. 
Harmon and Nolan combined both radar and earlier IR data to obtain 
a solution for Encke's shape and size giving a mean effective radius of 
2.42 km and an unusually large axial ratio of 2.6. 
Finally, comet Encke is also one of the very few comets to have its dust trail
imaged at visual wavelengths (Lowry et al. 2003; Sarugaku et al. 2005). 

Here we present results from new CCD observations of comet Encke, acquired at Steward Observatory's 
2.3m Bok Telescope on Kitt Peak. The data includes the first
detailed time series multi-color measurements of a cometary nucleus.
The observations were carried out in October 2002 when the comet 
was near aphelion at a heliocentric distance of 3.9 AU, and modest phase angle
of $\sim 7.2^{\circ}$. This epoch was close to some of the 
Fern\'{a}ndez et al. (2005) measurements, and so we attempt to link the two data sets to refine
our spin-rate result. In section \ref{Observations} we describe our observations in detail, 
as well as the lightcurve extraction techniques that were used. 
In section \ref{Color_variations} we investigate the time-series color measurements. This is 
followed by a discussion of the derived size and projected shape of the 
Encke nucleus in section \ref{Size_and_shape}.
Section \ref{rotation_rate} includes a detailed discussion of the rotation properties that 
have been extracted from the photometry, and our efforts to link them with the 
Fern\'{a}ndez et al. measurements, taken close in time to our observations. A summary of our 
main conclusions is presented in section \ref{SUMMARY}.

%%%%%%%%%%%%%%%%%%%%%%%%%%%%%%%%%%%%%%%%%%%%5
\noindent
\section{\bf Observations and lightcurve extraction}
\label{Observations}

CCD imaging of comet 2P/Encke was obtained using Steward
Observatory's 2.3m Bok telescope on Kitt Peak, on the nights of
October 3 and 4, 2002 (UT). The images were obtained using an 
NSF Lick 3 2048$\times$2048 pixel CCD at the telescope's Cassegrain focus. 
The pixel scale was 
0.30$^{\prime\prime}$/pixel in 2$\times$2 binned mode, and the total field 
of view was $\sim 5.1\times5.1$ arcmin. We used the Harris
$B$, $V$ and $R$-filters which have peak-transmission wavelengths at approximately
440 nm, 525 nm, and 590 nm, respectively. These filters are roughly
equivalent to the Johnson ($B$, $V$) and Kron-Cousins ($R$) filters.

Our aim was to obtain time-series $R$-filter photometry to determine 
the rotational properties of the nucleus from the lightcurve, and to obtain additional
time-series imaging in the $V$ and $B$ filters in order to assess the colors
of the object and to look for possible color variations as the object 
rotated. The observing strategy was to cycle through the three broadband 
filters by repeating the following sequence: $R-V-R-B-R-V-R-B-R$...
while stopping occasionally for standard star observations when 
the object passed across or near background stars. This sequence gave 
approximately twice as many $R$-filter images as $V$ and $B$, and thus denser
temporal/phase coverage of the lightcurve for the rotational analysis.

The night-1 observations (October 3) include 
18$\times R$-filter images with exposure times of 400 seconds, 
8$\times V$-filter images with exposure times of 400 seconds, and
7$\times B$-filter images with slightly longer exposure times of 500 seconds due to
the expected drop in S/N at this wavelength.
The night-2 observations (October 4) include 
20$\times R$-filter images, 
9$\times V$-filter images, and
8$\times B$-filter images, all with exposure times of 400-500 seconds.
For image processing purposes, a set of twilight sky
exposures were taken through each filter in use, and used to flat-field
the bias-subtracted images. All other instrumental artifacts such as
cosmic rays and bad rows/columns were removed in the standard manner. 

For each wavelength pass-band, the cometary instrumental
magnitudes were measured and compared with the brightnesses of several
non-variable background stars to extract the rotation lightcurve. 
This was applied to the $B$, $V$ and $R$-filter data. 
As all cometary images were taken with the telescope
tracking at the comet's apparent rate of motion, a slight trailing effect was
introduced to the background stars. However, the affect on the 
S/N of the lightcurve data points was negligible. 
These relative instrumental magnitudes were calibrated using observations 
of the Landolt standard field PG2213-006 (Landolt, 1992), taken at a range 
of airmasses throughout each night. Observing conditions were photometric
on both nights.

\begin{figure}[t]
\resizebox{\hsize}{!}
{
\includegraphics{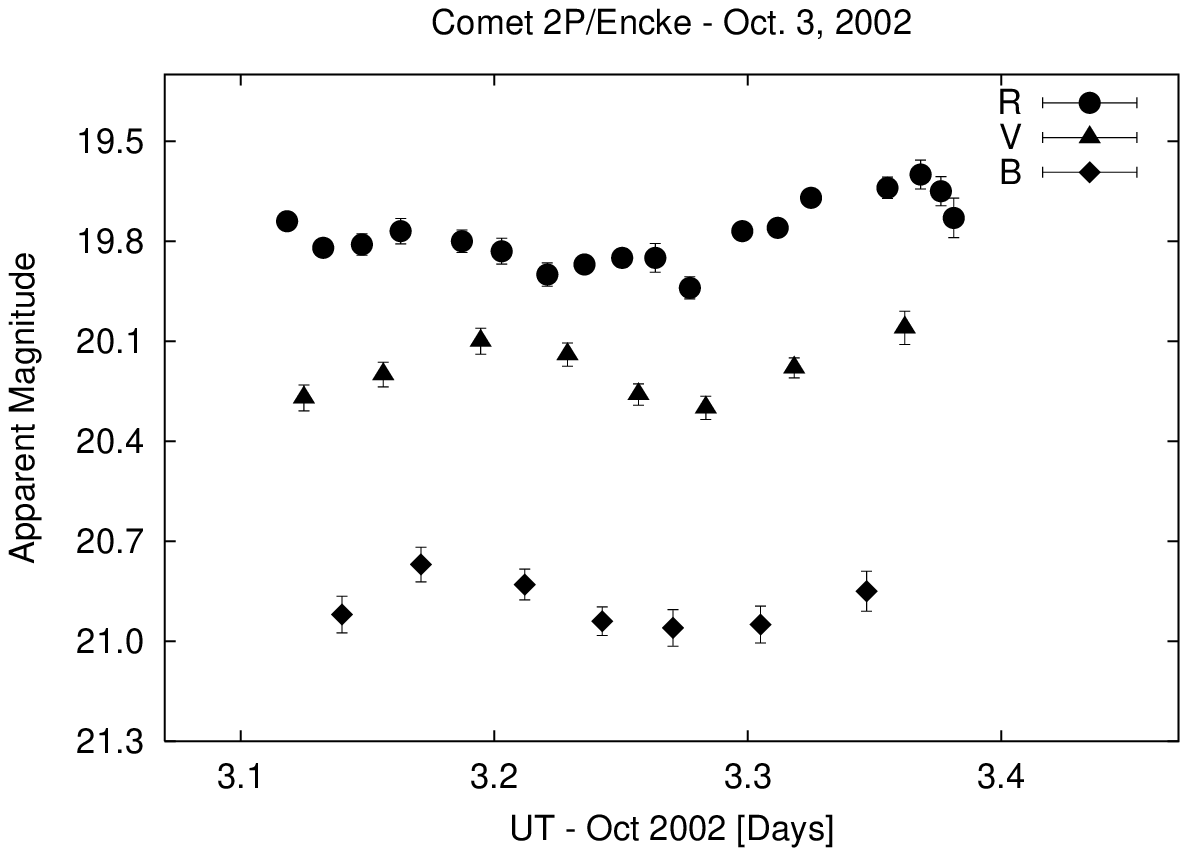}
\includegraphics{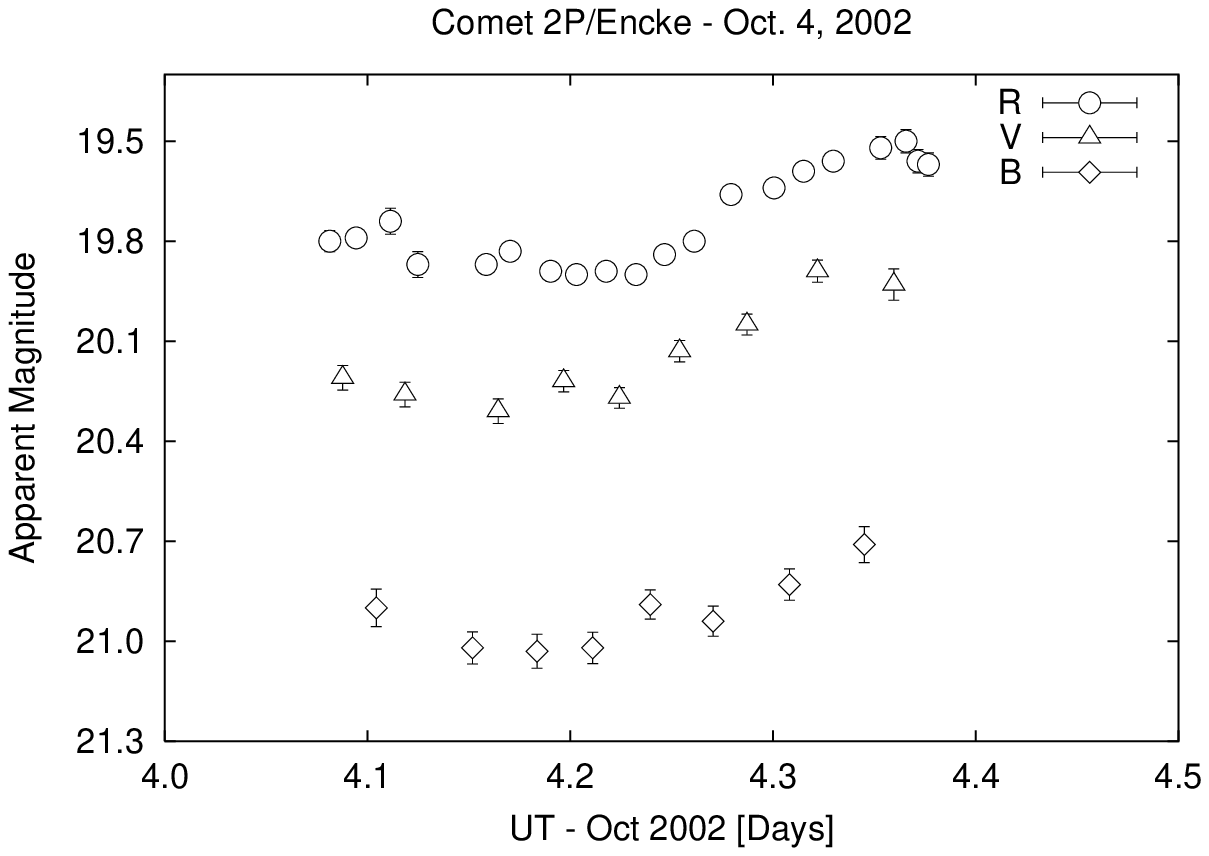}
}
\caption[]
{Time series $B$, $V$ and $R$-filter photometry for October 3 and 4, 2002.}
\label{fig_BVR_N12}
\end{figure}

Aperture photometry was performed on the comet 
and background stars through apertures large enough to take in at least 99\%
of the flux (i.e. 3$\times$FWHM of background PSF), therefore seeing changes between images had no effect on the
extracted lightcurve. The extracted $B$, $V$ and $R$-filter lightcurves are shown graphically in 
Figure \ref{fig_BVR_N12} for October 3 and 4. The $BVR$-filter apparent 
magnitudes and corresponding mid-exposure UT-Day are listed in 
Table \ref{Relative_mags_table}. The mean $B$, $V$ and $R$-filter apparent 
magnitudes are listed in Table \ref{Sizes_table} for each of the two nights 
of observation.
All image processing, photometry, and calibration were performed using the
IRAF program (Tody 1986, 1993).

\begin{figure}[t]
\resizebox{\hsize}{!}
{
\includegraphics{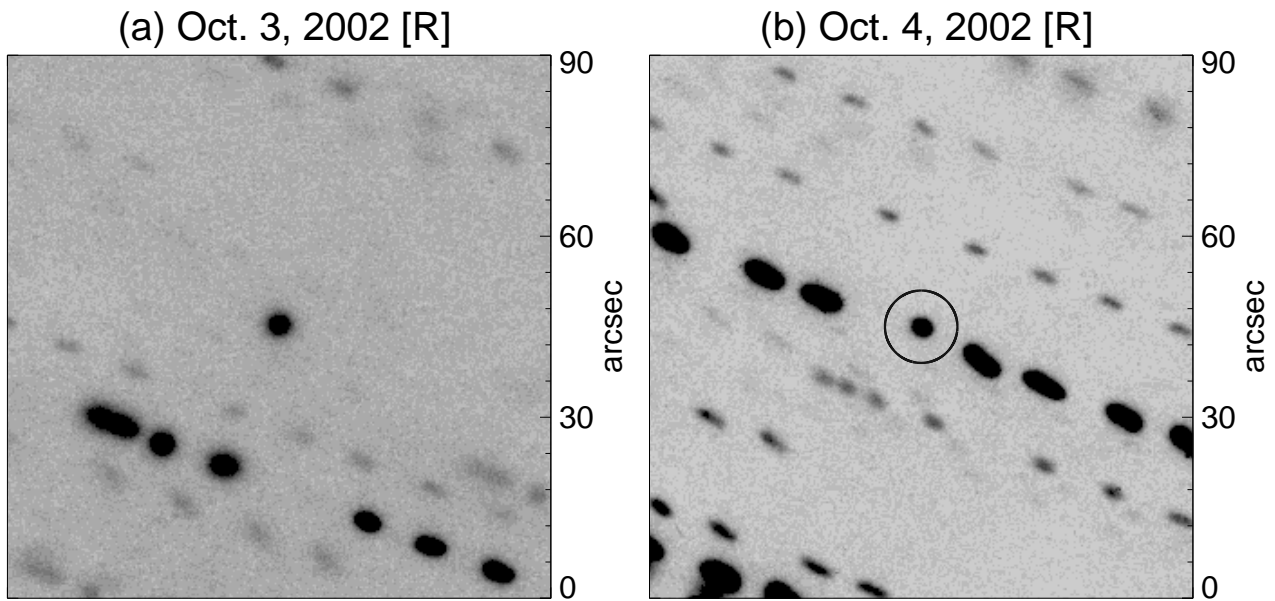}
}
\caption[]
{Co-added $R$-filter images of comet Encke from each 
night of observation, taken at Steward Observatory's 
2.3m Bok Telescope on Kitt Peak.  The comet is marked with a circle in panel (b).
The comet was stellar in appearance on both nights and in all filters.
}
\label{fig_2P_CCDimages}
\end{figure}

We searched for signs of a resolvable coma around the comet by measuring the 
surface brightness profile of the comet and comparing with the stellar-background 
brightness profile (or PSF). As all cometary images were taken with the telescope
tracking at the comet's apparent rate of motion, the cometary profile was measured by aligning
all $R$-filter frames on a given night on the comet and co-adding them, and
then measuring the azimuthally averaged brightness profile through a series of circular
annulli. As the background stars were slightly trailed in each exposure, the 
brightness profile was extracted using the method of Lowry and Fitzsimmons (2005). 
This method provides the effective brightness profile which would have been 
obtained under sidereal tracking. The profiles were indistinguishable. 
In Figure \ref{fig_2P_CCDimages} we show the $R$-filter co-added images from each night, and one 
can see that there is no evidence of a tail, or even the dust trail that was
previously detected at optical wavelengths 
(Lowry et al. 2003; Sarugaku et al. 2005). 
The comet appears as a sharp point source.

%%%%%%%%%%%%%%%%%%%%%%%%%%%%%%%%%%%%%%%%%%%%%%%%%%%%%%%%%%%%%%%%%%%%%%%
\noindent
\section{\bf Color variations}
\label{Color_variations}

\begin{figure}[t]
\resizebox{\hsize}{!}
{
\includegraphics{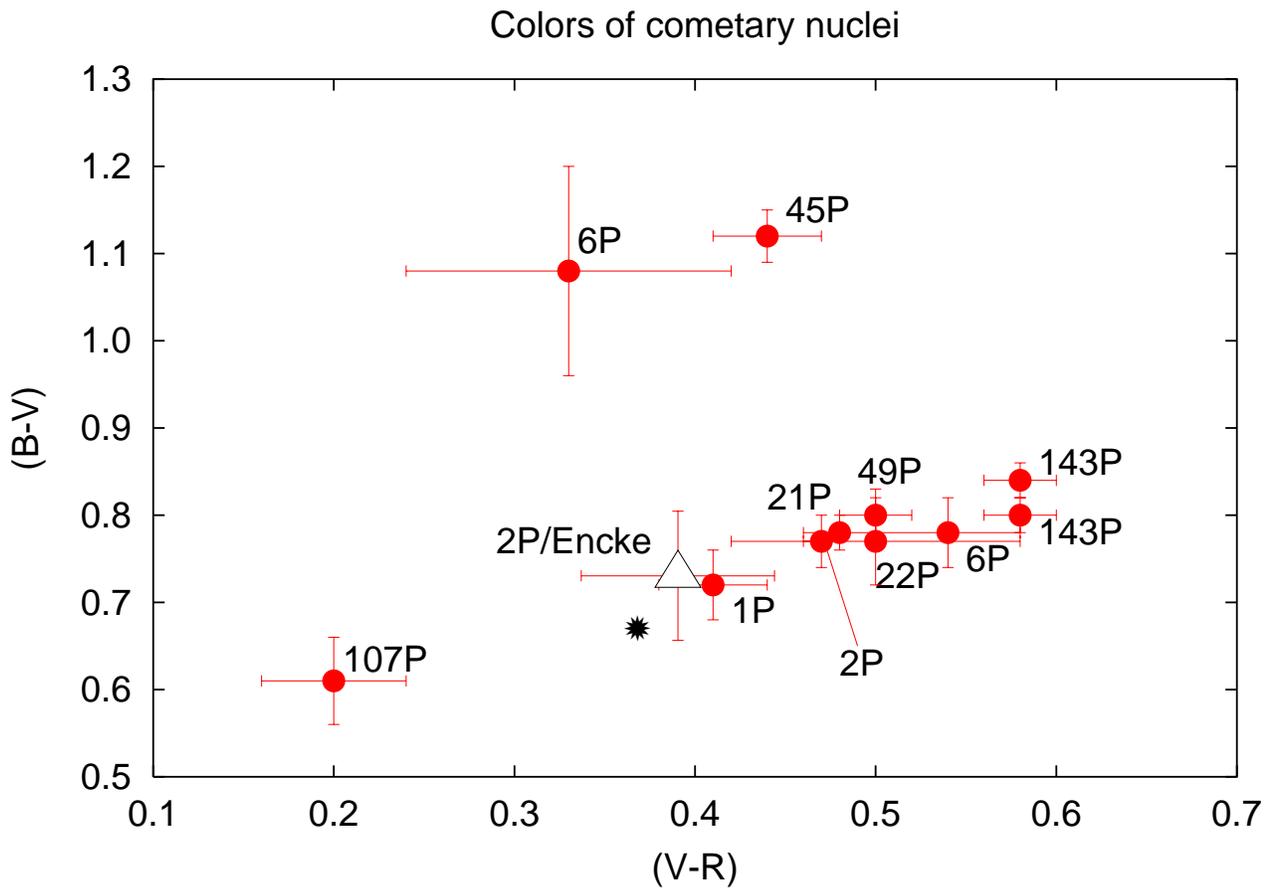}
}
\caption[]
{Comparison of our colors for comet 2P/Encke with those of other Jupiter-family 
nuclei for which $BVR$ measurements were obtained together. Encke's colors are 
slightly redder than the equivalent Solar values, but are bluer than most other
comets as well as previous values measured for Encke. Color data taken from review
by Lamy et al. (2005). Solar colours are marked with a star.}
\label{fig_colour_comp}
\end{figure}

The imaging sequence discussed above was performed so as to simplify, as much as
possible, the interplotation between apparent magnitudes from the three filters for the
color determination. Without adequate interpolation between filters, the resulting color measurements
will be confounded by the effects of varying nucleus cross-section as the object rotates. 
Each $V$-filter image was bracketed by two $R$-filter images. Therefore, to obtain the ($V$-$R$)
color index at the time of the $V$-filter image, we took the average of the two adjacent $R$-filter apparent 
magnitudes and subtracted this from the $V$-filter apparent magnitude value.
As for the ($B$-$V$), each $B$-filter image was bracketed by two $V$ and $R$-filter images in 
the following way: $V-R-B-R-V$, therefore we took to the average of the two nearest $V$-filter apparent
magnitudes and subtracted the average from the $B$-filter apparent magnitude value. 
Table \ref{Relative_mags_table} lists the color values obtained.
The average color indices for October 3 are:
($V-R$) = $0.40 \pm 0.06$ and ($B-V$) = $0.70 \pm 0.08$.
The average color indices for October 4 are:
($V-R$) = $0.38 \pm 0.05$ and ($B-V$) = $0.76 \pm 0.07$.
Encke's colors are slightly redder than the equivalent Solar values of 
($V-R$)$_{\odot}$ = 0.36, and ($B-V$)$_{\odot}$ = 0.67, but are bluer than most 
other comets as well as previous values for Encke (Luu and Jewitt 1990). This is illustrated in 
Figure \ref{fig_colour_comp} where we compare our colors with those comets for which $BVR$
photometry exists.

\begin{figure}[t]
\resizebox{\hsize}{!}
{
\includegraphics{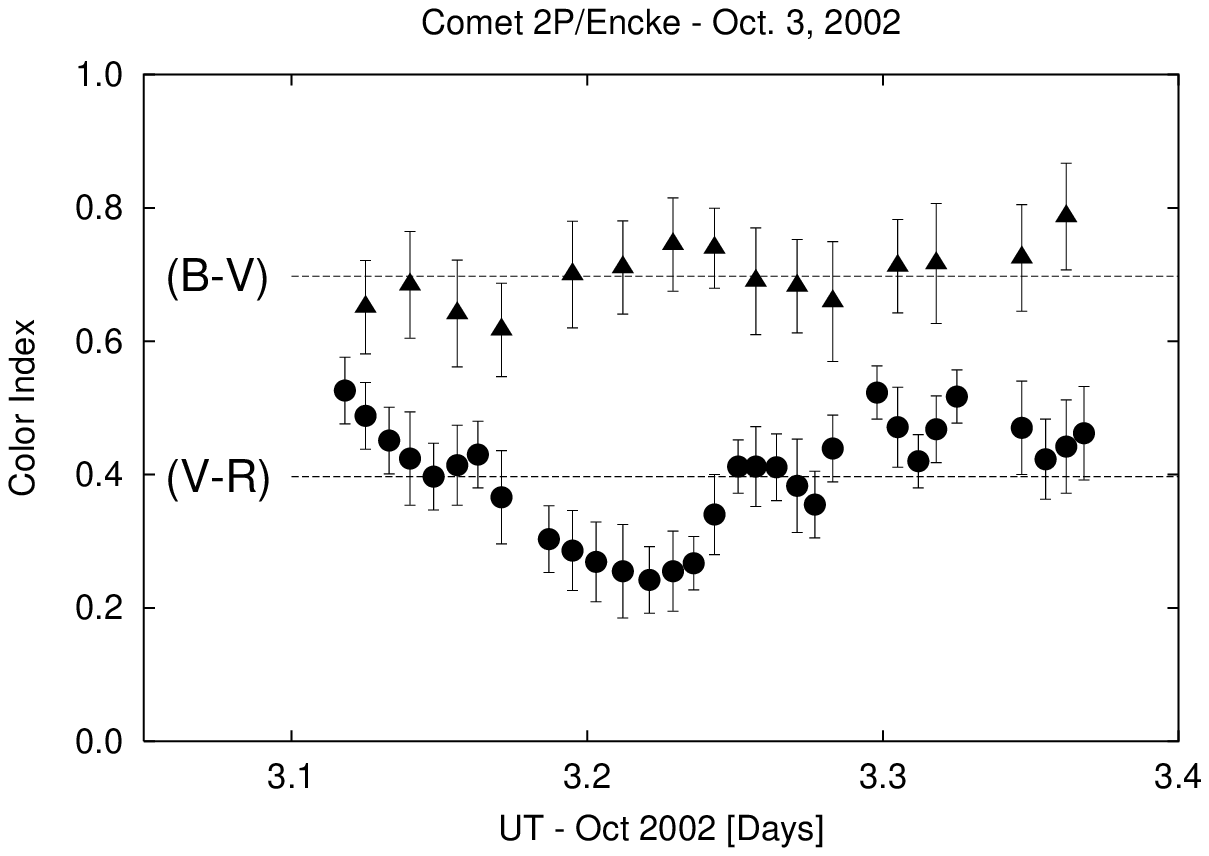}
\includegraphics{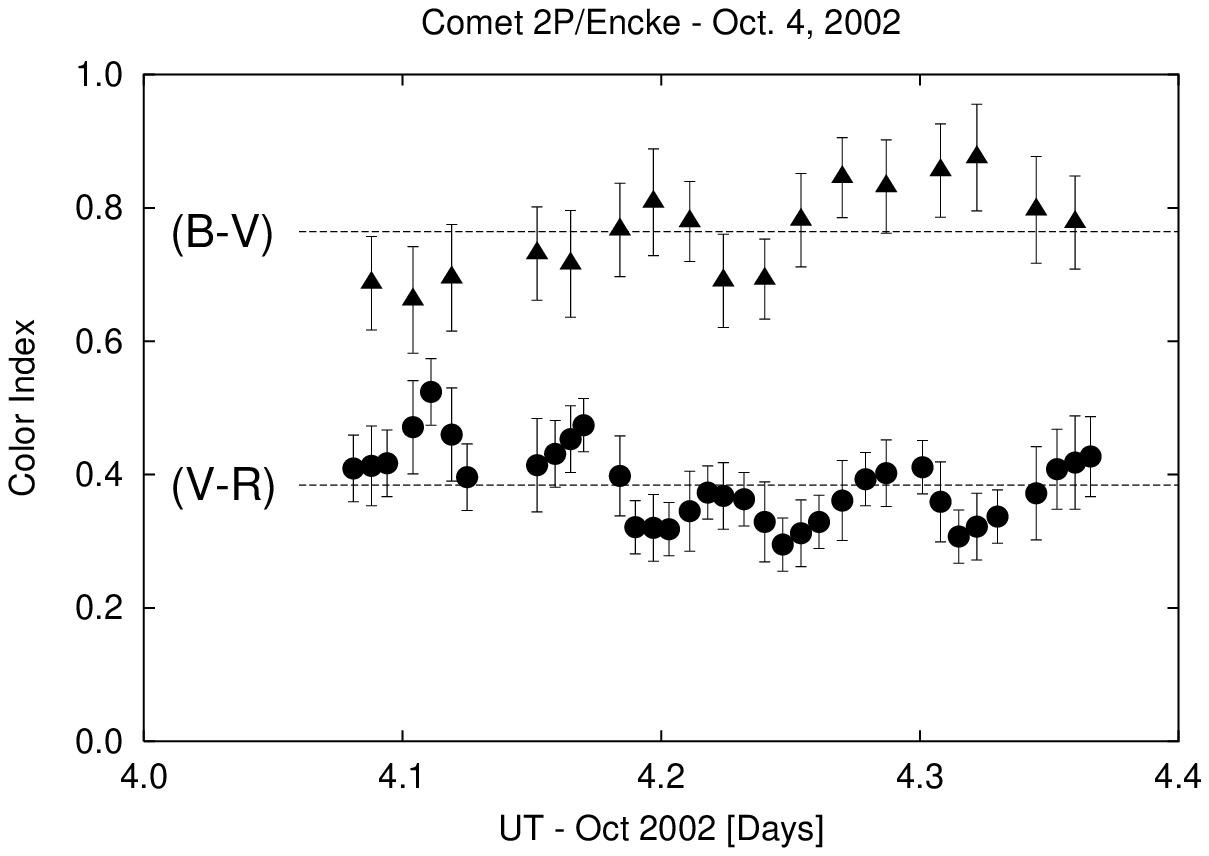}
}
\caption[]
{Color indices versus time for each night of observation. No significant color variations are seen for
night 2, but the beginning of night 1 shows a clear and systematic dip in ($V-R$) color,
which is not seen for ($B-V$).
This may be due to either imhomogenious nucleus surface properties or a brief outburst in
gaseous emissions visible at $B$ and $V$ wavelengths around Oct. 3.2 UT.
}
\label{fig_2P_oct02_colours_N12}
\end{figure}

We searched for possible variations of color with rotation by 
plotting the $BVR$ color indices versus UT-Day and the results are shown in 
Figure \ref{fig_2P_oct02_colours_N12}.
No significant color variations are seen for October 4, but on October 3 we see
a clear and systematic dip in ($V-R$) color index, which is not seen for ($B-V$).
This systematic variation is considered real at the 2$\sigma$ confidence level.
Also, there was no sign of time-varying atmospheric extinction, from inspection of the
brightnesses of the background stars.
However, if we are looking at a bare nucleus and the 11.089 hour rotation period
from section \ref{Combining_with_sep2002}
is the correct one - and we strongly suspect that it is - then color variations
should be repeated from night to night. 

We explore the possibility that a brief outburst in
gaseous emission occurred around Oct. 3.2 UT.
C$_{2}$ molecules within cometary coma have prominent emission bands 
at wavelengths that overlap the passbands of the $B$ and $V$ broadband filters. 
Brief outbursts in activity were recently detected by the Deep Impact 
probe on approach to comet 9P/Tempel 1 (A'Hearn et al. 2005). These outburst were 
due to active areas coming into view of the sun as the comet rotated. 
For some of the larger outbursts, residual activity was seen for up to 18 hours
after the time of peak brightness. In the case of Encke, such an outburst would 
need to be a singular, transient event, not repeated on subsequent rotations of 
the nucleus as the
active spot again comes into sunlight, at least during the two nights of
our observations.
This would satisfy the criterion that if 
the $\sim$11 hour period is correct then we should be looking at roughly the 
same hemisphere of the nucleus on October 3 and 4 and thus the
color variations would be the same from night to night for a 
bare nucleus. 
If these color variations were due to an impulsive release of gaseous
species then as the color-change event lasted approximately
2.4~hrs, the gas would need to completely traverse our photometric 
aperture within this time. At a geocentric distance of 3.02~AU,
the gas expansion velocities would need to be at least 1.2 km~s$^{-1}$
for our chosen aperture sizes. This is at least a factor of $\sim~3-4$
times larger than the expected expansion velocities for these species at a 
heliocentric distance of 3.93~AU, where the expected blackbody sub-Solar
temperature on the Encke nucleus would be $\sim$~204~K. Therefore, this scenario is very unlikely.
Also, the active-comet scenario is further weakened as no coma or tail was 
seen in our deep images of the comet, nor in previous near-aphelion imaging. 

We can conclude that the color variations are significant and most likely 
imply inhomogeneity on the surface of Encke's nucleus 
(or be due to a combination of nucleus-surface and coma properties). 
Whether there is coma present
or not, we do demonstrate that color variations of this order
are possible to detect. 
Complex rotation, in conjunction with a highly elongated nucleus with 
significant surface inhomogenieties, may well explain both the color variations
and the erratic photometric behaviour of the comet when it is near aphelion
(see section \ref{Size_and_shape}).

As a note for future comet-nucleus observers - we feel that it is important 
to focus on $R$-filter 
photometry for determining the rotation period from time-series photometry,
to avoid these emission bands as coma may be present.
Had the $V$ or $B$ filter been the only filter used, then our period analysis
may have been adversely affected.
We encourage future lightcurve investigations to 
include full rotational phase coverage at multiple bandpasses, to search 
for possible color variations on cometary nucleus surfaces, and to give 
added dimension to the rotation period analysis. 

%%%%%%%%%%%%%%%%%%%%%%%%%%%%%%%%%%%%%%%%%%%%%%%%%%%%%%%%%%%%%%%%%%%%%%%

\noindent
\section{\bf Nucleus size and shape}
\label{Size_and_shape}

We calculated the mean nucleus radius in each of three filters for both nights using,
\begin{equation}
\label{Mag_Size}
A_{\lambda} r_{N}^{2} = 2.238 \times 10^{22} R_{h}^{2} \Delta^{2} 10^{0.4(m_{\odot \lambda} - m_{\lambda} + \alpha \beta)} 
\end{equation}
where $m_{\lambda}$ is the average $B$, $V$ or $R$ apparent magnitude for a given night,
$A_{\lambda}$ is the geometric albedo in the respective filter, $r_{N}$ [m] 
is the nucleus effective radius, $R_{h}$ [AU] and $\Delta$ [AU] are 
the heliocentric and geocentric distances, respectively, 
$\alpha$ and $\beta$ are the phase angle and phase coefficient, respectively, 
and $m_{\odot \lambda}$  is the apparent $B$, $V$ or $R$ magnitude of the Sun. For
these calculations we use 
$m_{\odot R}$ = -27.26, ($V-R$)$_{\odot}$ = 0.36, and ($B-V$)$_{\odot}$ = 0.67.
The derived mean radii for each filter and each night are listed in Table \ref{Sizes_table}.
We use the empirically derived values for the geometric
albedo and phase coefficient of 0.047 and 0.06 magnitudes/degree, respectively
(Fern\'{a}ndez et al. 2000). 
The average $R$-filter apparent magnitude across both nights is
$19.76 \pm 0.03$, which corresponds to a mean effective radius of
$3.95 \pm 0.06$ km. 

We can scale this average $R$-filter apparent magnitude to the geometry
of the September 2002 photometry listed in Fern\'{a}ndez et al. (2005), using 
the standard $5log(R_h \Delta)$ scaling law for an inert body,
along with the linear phase-darkening correction using the phase coefficient above. 
We find that the scaled October magnitude is $19.40 \pm 0.03$ which is very close to the 
Fern\'{a}ndez et al. average $R$-magnitude of $19.34 \pm 0.02$ 
(or mean radius = $4.06 \pm 0.04$ km), and 
agrees at the 2$\sigma$ level, implying an inactive nucleus on both datasets. 
A 1$\sigma$ agreement
can be achieved with an increase in the phase coefficient of just $0.01$. 
Alternatively, if one assumes that the nucleus is active and it has a typical dust-production 
dependence on heliocentric distance (i.e. the brightness would scale as
$10log(R_h)$), then the scaled October magnitude is at least 0.1 magnitudes 
too faint for agreement with the Fern\'{a}ndez et al. average $R$-magnitude.
Of course, the differences in heliocentric and geocentric distances 
between the two data sets are small and so scaling of this type cannot definitively 
rule out the presence of an unresolved coma. 
The mean-radius 
for the other data sets listed in Fernandez et al., taken much farther back in
time than our October 2002 measurements, are as follows: 
July 19, 2001 - $4.19 \pm 0.05$ km;
August 10-13, 2001 - $3.24 \pm 0.04$ km;
September 21-25, 2001 - $4.61 \pm 0.05$ km; and 
October 6-8, 2001 - $2.36 \pm 0.09$ km.
Again, these values assume a common albedo and phase coefficient.

We now compare our mean $R$-magnitude with other
published data that include a good sampling of the $R$-filter lightcurve. 
The photometry of
Jewitt and Meech (1987) results in a mean-radius of $3.75 \pm 0.09$ km, 
using their 1985 data set and using the empirically derived values 
for the albedo and phase coefficient. This value is close to ours of
$3.95 \pm 0.06$ km.
However, the time-series photometry from Luu and Jewitt (1990) imply 
a smaller mean radius of $3.28 \pm 0.06$ km. 
In three of the eight cases where $R$-filter 
lightcurves were obtained - and for the radar+IR study from Nolan and Harmon (2005) 
noted in section \ref{INTRODUCTION} - the mean radius was measured
to be between $2.36$ and $3.28$ km. For the remaining five cases,
the mean radius falls in the range
of $3.75-4.61$ km. Although there is good agreement between most of the data
sets, clearly the photometric behaviour of comet Encke is still perplexing, 
and may imply an erratically behaved unresolved coma with a very steep 
brightness profile 
(as the flux contribution from a shallow steady-state coma is only a few percent of 
the total measured flux), and/or complex rotation of a very elongated nucleus. 
There is no correlation between the observed mean brightness and the observing 
geometry or orbital position of the available data sets. 
This erratic behaviour, along with the color variations noted above, 
will require further data and complex modelling to fully understand
(the latter being outside the scope of this particular paper).

The observed magnitude range $\Delta m$
was $0.35 \pm 0.05$ and $0.40 \pm 0.04$ for October 3 and 4,
respectively. This result uses the $R$-filter data only as the lightcurve coverage
was best for this wavelength passband. We can infer limits on the nucleus shape using:
\begin{equation}
\label{Axis_ratio}
\frac{a}{b} \geq 10^{0.4 \Delta m}
\end{equation}
where $a$, $b$ are the semi-axes of the nucleus. 
This simple model assumes a bi-axial ellipsoid shape 
and uniform surface albedo, and ignores phase effects. Taking the maximum
observed brightness range, i.e. from October 4, we get 
$a/b \geq 1.44 \pm 0.06$ for the Encke nucleus. Applying this axial ratio to
our $R$-filter photometry gives nucleus semi-axes of 
[$3.60 \pm 0.09$]$\times$[$5.20 \pm 0.13$] km. These dimensions, as with our
measured mean radius above, could well be upper limits to the size, given the
erratic behaviour of the photometry across the various published data sets, possibly
implying outgassing. Although again, we point out that no resolved coma or tail
has been seen in any of the above data sets, even with very deep-imaging 
techniques applied in most cases.

%%%%%%%%%%%%%%%%%%%%%%%%%%%%%%%%%%%%%%%%%%%%%%%%%%%%%%%%%%%%%%%%%%%%%%%%

\noindent
\section{\bf Nucleus rotation rate}
\label{rotation_rate}

We applied the method of Harris et al.~(1989) 
to determine the rotation period from the time-series photometry. We focus on the $R$-filter
data to constrain the rotation period as we have roughly twice as many data points
at this passband.
This method involves fitting an $n$th-order Fourier series to the
relative magnitudes, which is then repeated
for a wide range of periods until the fit residuals are 
minimized. We created periodograms over the reasonable range of 
$\sim$ 1--30 hours. This range includes all previously proposed periods for 
Encke noted in section \ref{INTRODUCTION}.
When fitting model lightcurves to the data, we start off with low order fits 
to get a more accurate feel for where the prominent periodicities reside. 
We then increase the fit
order to refine the dominant periodicity and its associated uncertainty, and
also the quality of the fit. We stop increasing the order once  
the quality of the fit no longer improves. The dominant periodicity must be
consistent with our initial visual inspections.

\begin{figure}[t]
\resizebox{(\hsize)/2}{!}
{
\includegraphics{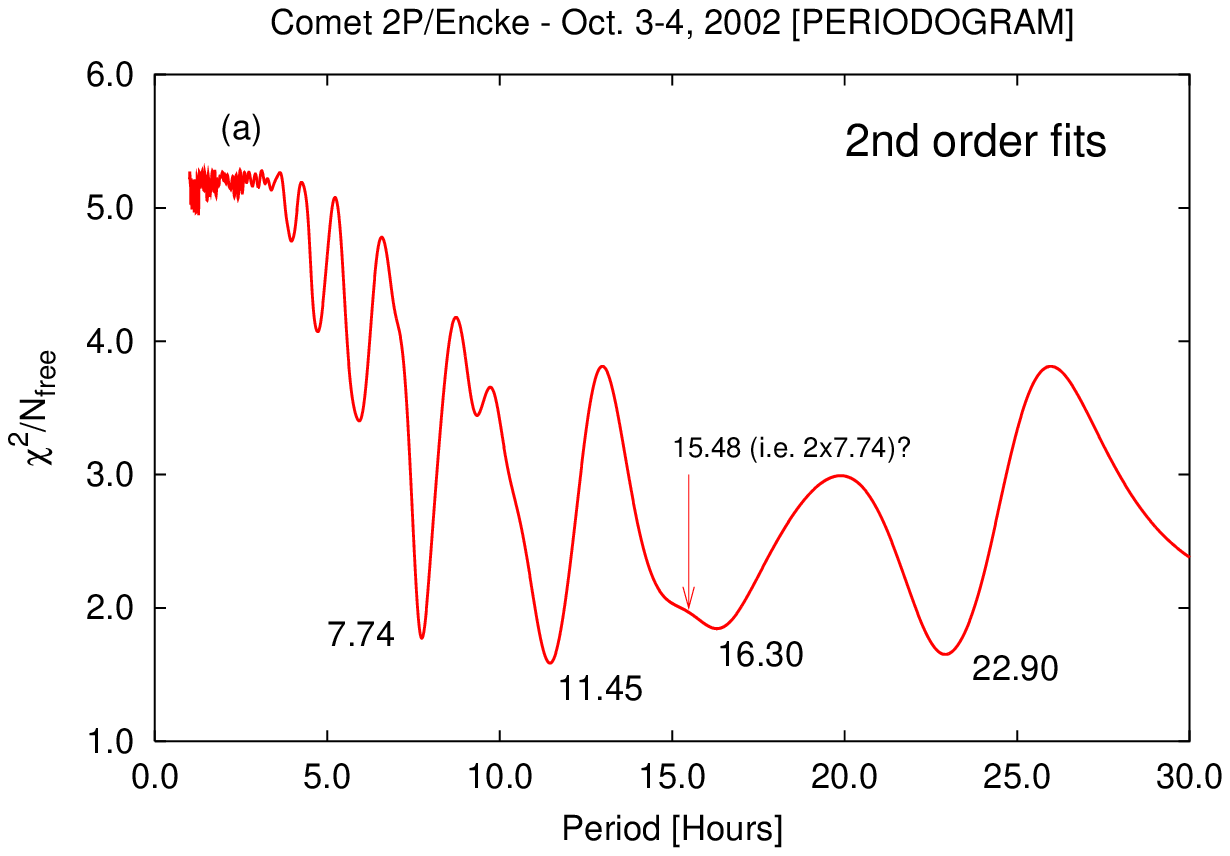}
}
\resizebox{(\hsize)/2}{!}
{
\includegraphics{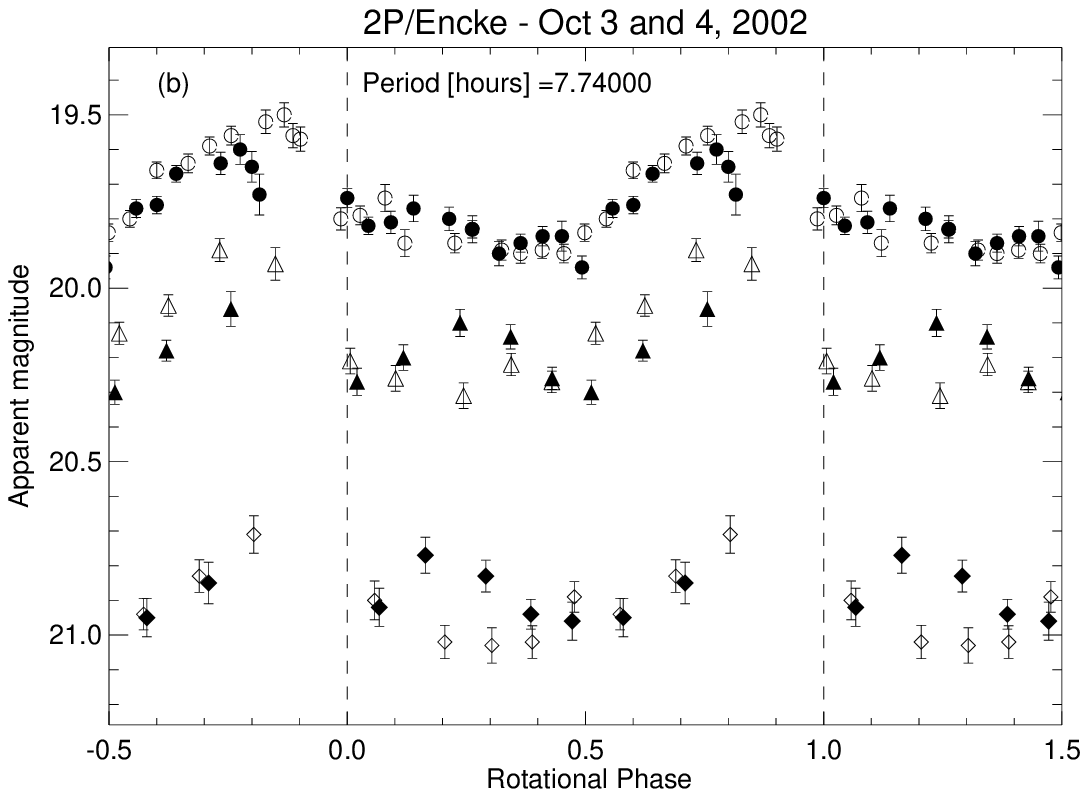}
}
\resizebox{(\hsize)/2}{!}
{
\includegraphics{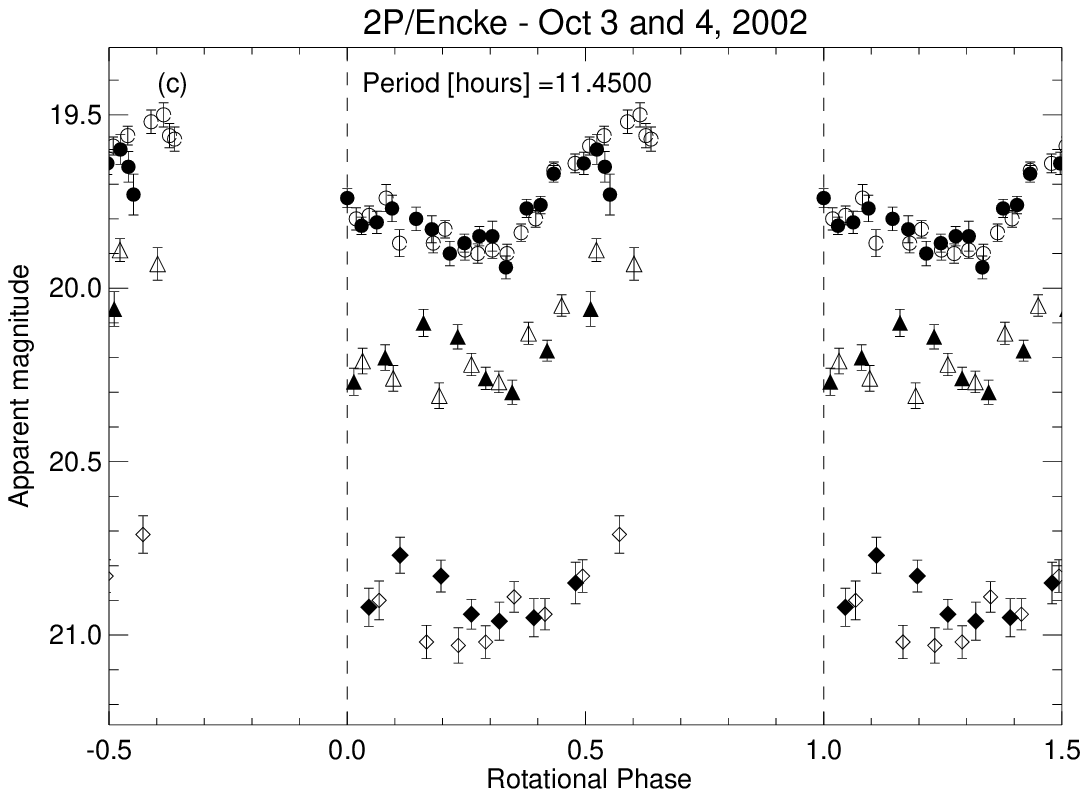}
}
\resizebox{(\hsize)/2}{!}
{
\includegraphics{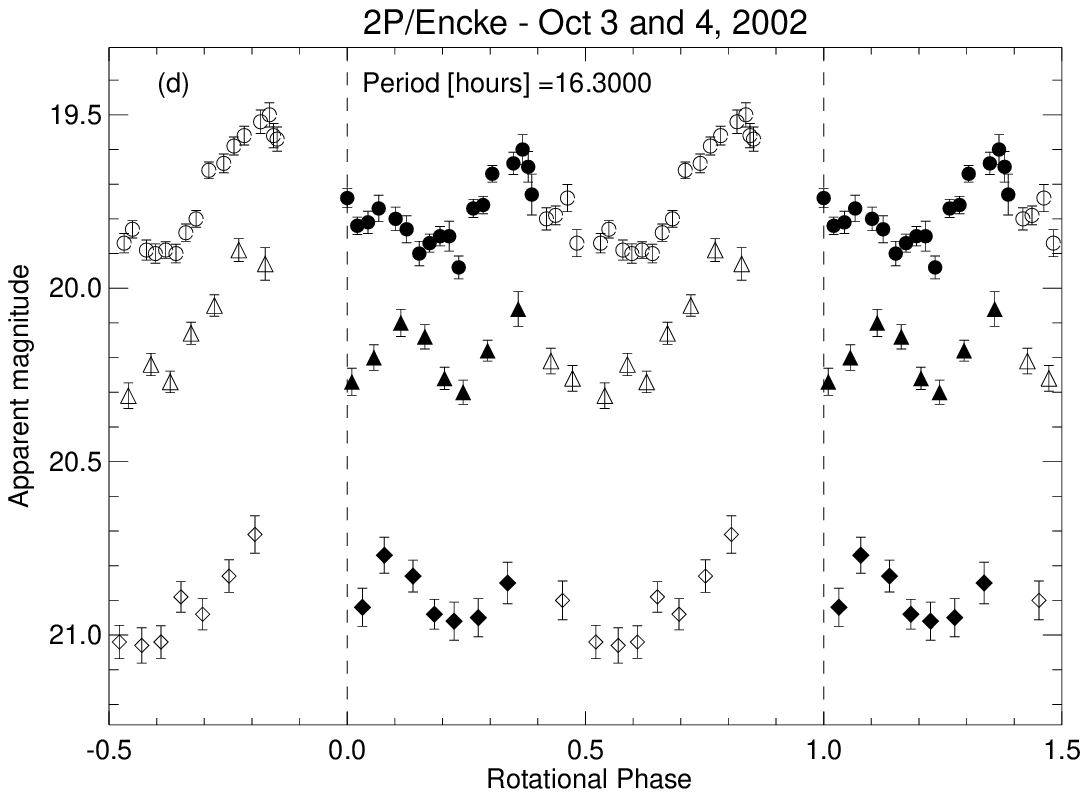}
}
\resizebox{(\hsize)/2}{!}
{
\includegraphics{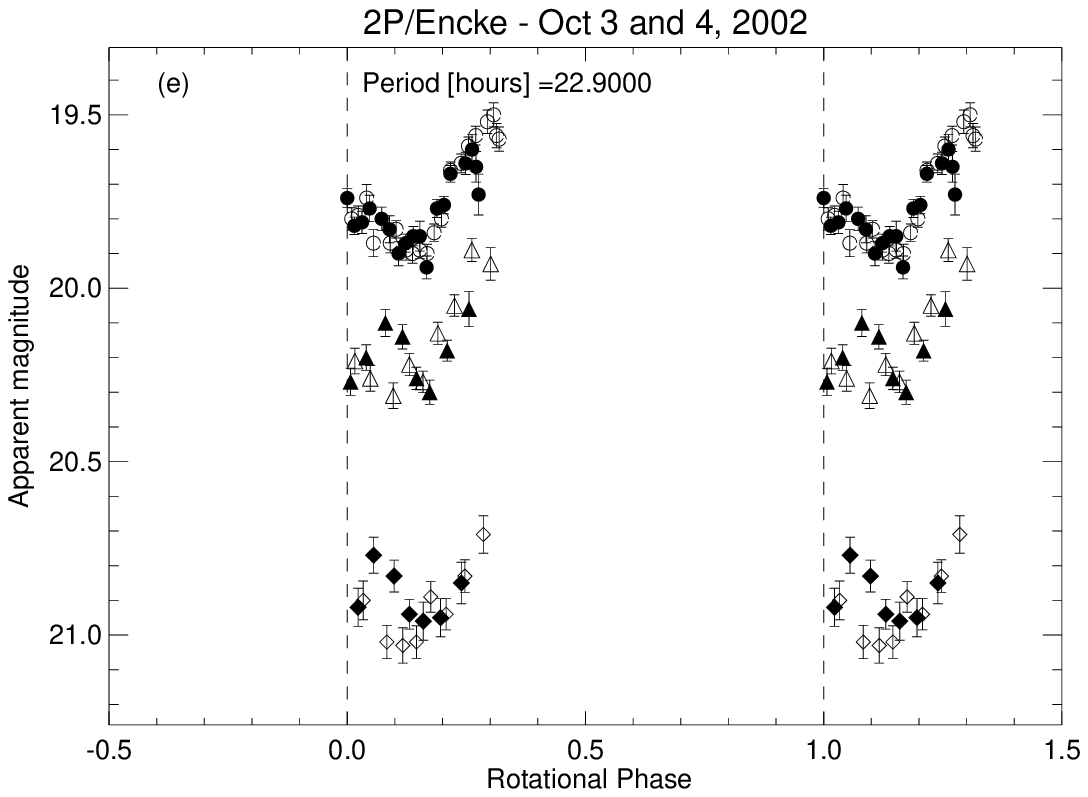}
}
\resizebox{(\hsize)/2}{!}
{
\includegraphics{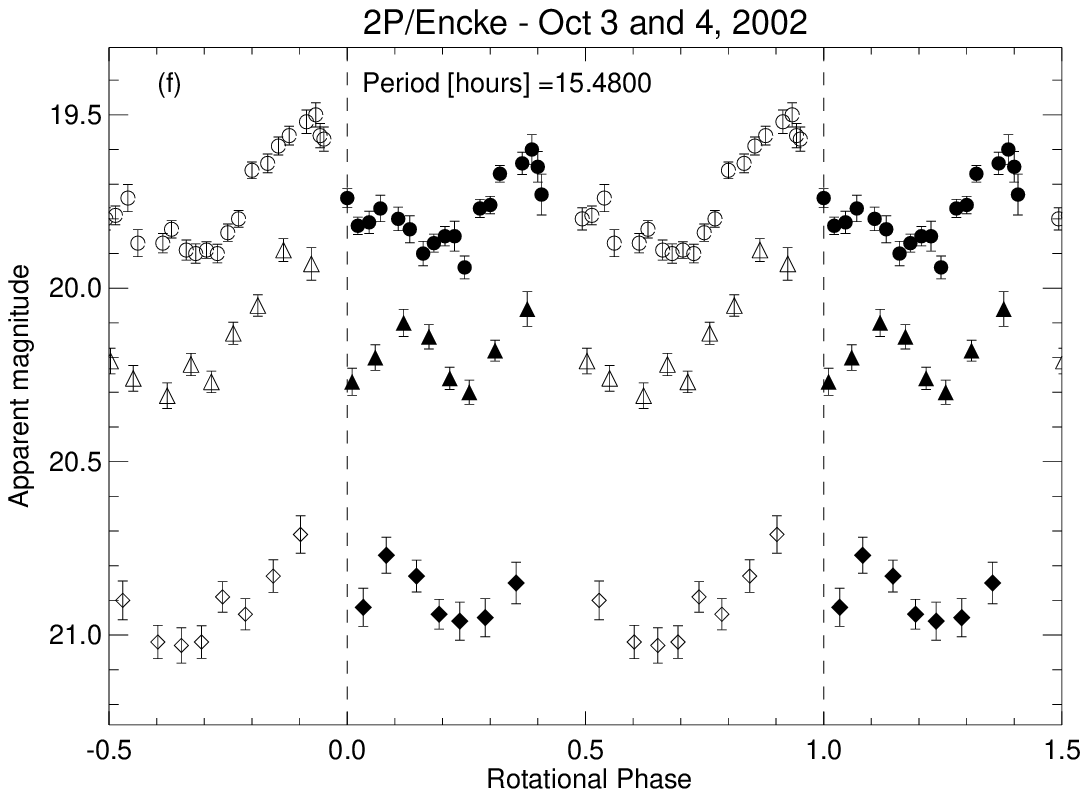}
}
\caption[]
{(a) Periodogram from 2nd order fourier fits to the $R$-filter data. The four most
prominent features are marked. (b-f) The $BVR$ magnitudes phased to each
of these four periodicities, along with a phasing at 15.48 hours (see text). 
We find that the 11.45 hour
period offers the best solution, but does not necessarily fit well across all three filters.
When we combine our $R$-filter data with the September 2002 data set from 
Fernandez et al. (2005), only a period near 11 hours offers a common solution
(see section \ref{Combining_with_sep2002}). Filled symbols - October 3rd; Open symbols -
October 4th. $R$-filter data are denoted by circles, $V$-filter by triangles, and $B$-filter
by squares.
}
\label{fig_oct_phase_plots}
\end{figure}

This method was applied to the $R$-filter data and 
the resultant periodogram from 2nd order fits is shown in 
Figure \ref{fig_oct_phase_plots} (a). There are four main features at 
7.74, 11.45, 16.30 and 22.90 hours, and the phased $R$-band light-curves look believable 
at all four periods [see Figures \ref{fig_oct_phase_plots} (b)-(e)]. 
The $11.45 \pm 0.11$ hour period produces the best fit, 
but this value is longer than previous determinations 
(Fernandez et al. 2005; Lowry et al. 2003). 
The measurement uncertainty is quite 
large, which is most likely due to reduced temporal coverage.
The period at 7.74 hours does not satisfy the $B$ and $V$-filter data, 
particularly $V$, given the very poor overlap of data points between the two nights.
Also, there is the obvious fact that it is a single-peaked lightcurve, which
is physically unrealistic. There is similarly poor overlap in the phased 
$B$ and $V$ data points for the 11.45 hour period (and 22.90 = 2$\times$11.45 hours), 
although to a lesser extent. These are attributable to the color differences
between the two nights, noted in section \ref{Color_variations}.

Finally, the wide periodogram feature at 16.30 hours requires some discussion.
We essentially rule out this period as it is most likely due to the period-fitting software
slotting the two lightcurves end-to-end, leading to a false periodicity. 
Of course, this does not mean that such periodicities are not viable solutions,
but in this case there are no prominent harmonic features associated with 
this period. Notice how wide the feature is. We speculate
that this may be a merging of two separate features at 16.30 hours and
perhaps one at 15.48 hours (i.e. 2 $\times$ 7.74 hours), where the curve seems
to level off slightly. Even though this is close to the previously published period at
15.08 hours (Luu and Jewitt 1990), and a could explain our 
observed color differences between Oct 3 and 4, we conclude that both periods
at 15.48 and 16.30 hours are likely pathological cases that appear to fit only because
there is no overlap in the phased data from the two nights.
Also, the persistence of the 11 hour period producing the best fit over most of the 
available data cannot be taken lightly as shown by Fern\'{a}ndez et al. (2005), 
and as we go on to show now when we link our data with some of the 
Fern\'{a}ndez et al. data set.

%%%%%%%%%%%%%%%%%%%%%%%%%%%%%%%%%%%%%%%%%%%%%%%%%%%%%%%%%%%%%%%%%%%%%%%%%

\subsection{\bf Combining our data with Sep 2002}
\label{Combining_with_sep2002}

The Fern\'{a}ndez et al. (2005) results are based on a very large data set of
$R$-filter magnitudes spanning July 2001 to September 2002. We consider here
the data from September 10-16, 2002 only as this was obtained close to our 
current data set taken just 2-3 weeks later. The Fern\'{a}ndez et al. September 2002
data were taken using the 
University of Hawaii's 2.2m telescope on Mauna Kea, and Steward Observatory's 
2.3m Bok telescope at Kitt Peak, and contains 296$\times R$ filter data points.
The comet was at heliocentric
and geocentric distances of 3.97 AU and 2.97 AU, and the phase angle was 
$\sim$ 1.6$^{\circ}$. 
Of course, if we combine the two data sets then the resultant period will be
naturally biased towards the Fern\'{a}ndez et al. value as 
most of the data are from that set. The main aim of this execise is 
to refine our period from section \ref{rotation_rate}
by combining our data with their September 2002 data, 
i.e. taken approximately two to three weeks earlier than ours, thus substantially 
increasing the time-base, and 
verifying that a period just over 11 hours really can satisfy both data sets.
The derived period must successfully phase the October data onto the
September data (i.e. the phased brightness maxima and minima must coincide). 

\begin{figure}[t]
\resizebox{(\hsize)/2}{!}
{
\includegraphics{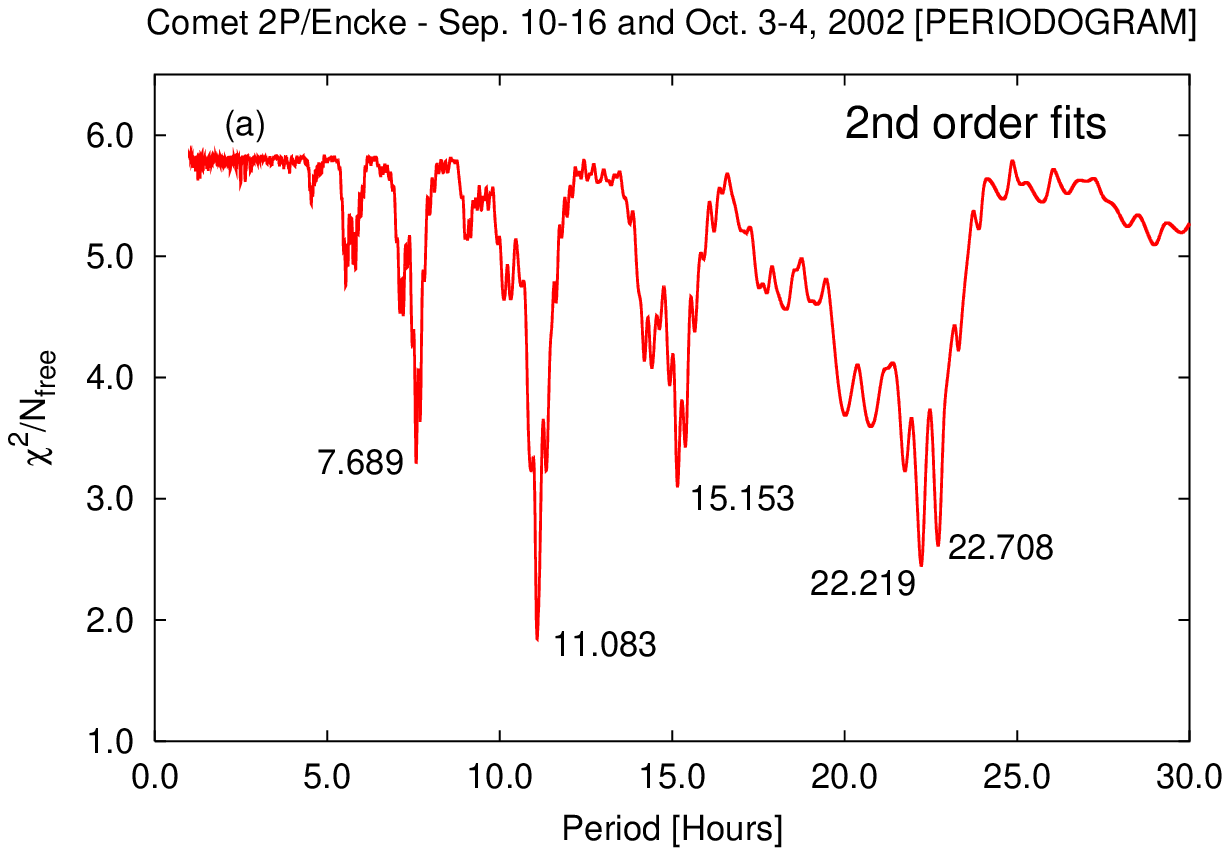}
}
\resizebox{(\hsize)/2}{!}
{
\includegraphics{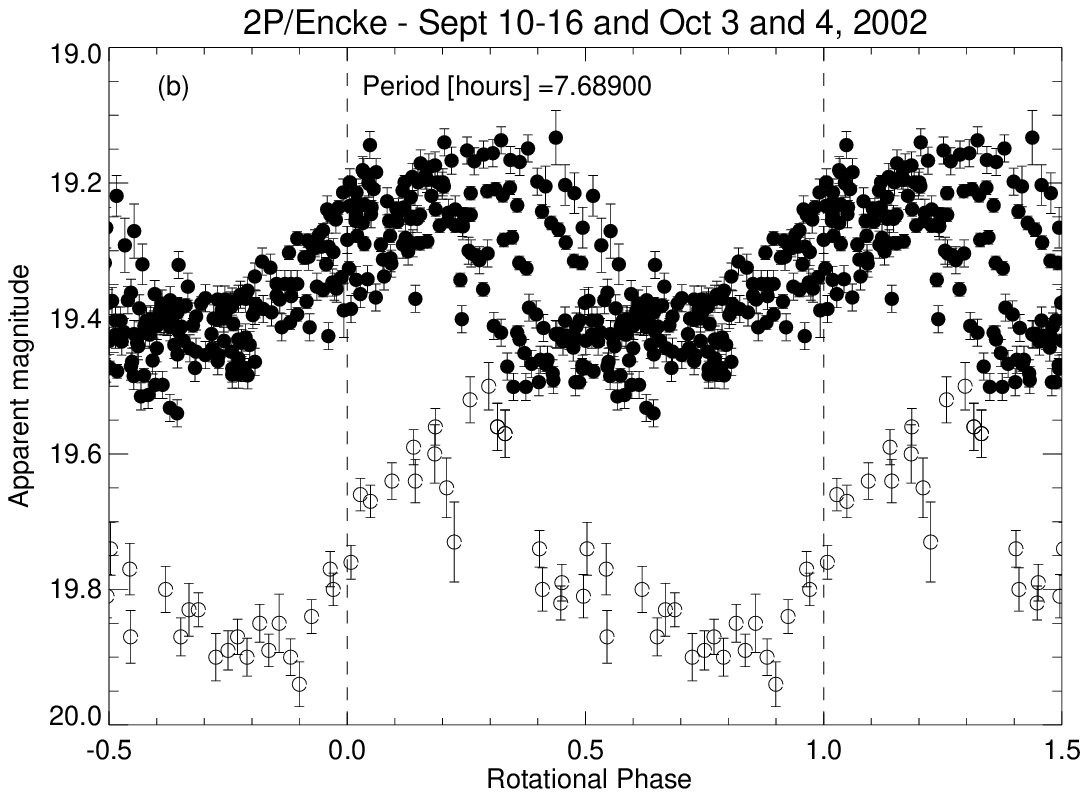}
}
\resizebox{(\hsize)/2}{!}
{
\includegraphics{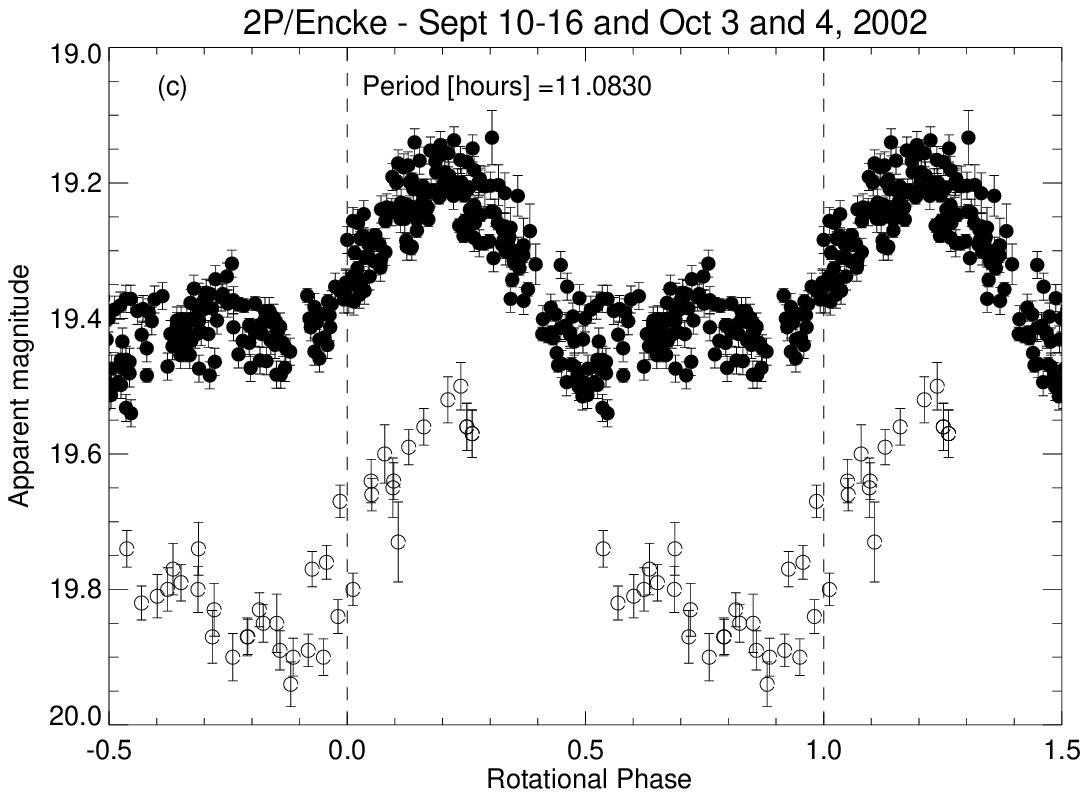}
}
\resizebox{(\hsize)/2}{!}
{
\includegraphics{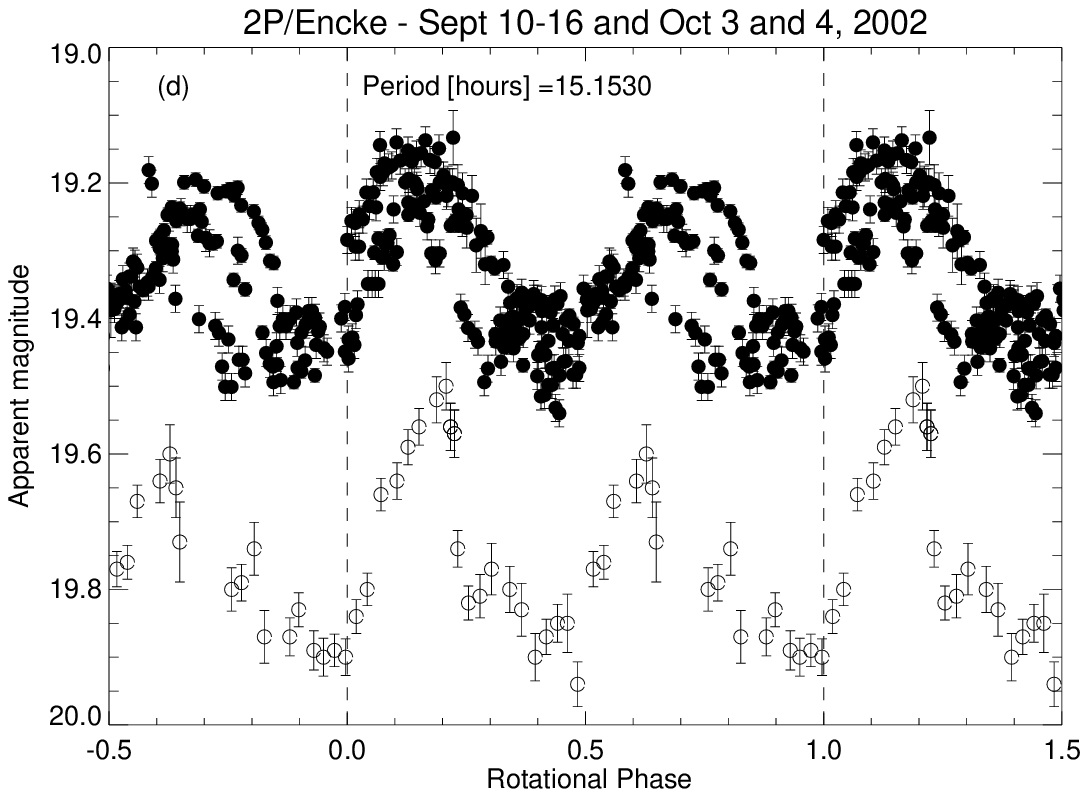}
}
\resizebox{(\hsize)/2}{!}
{
\includegraphics{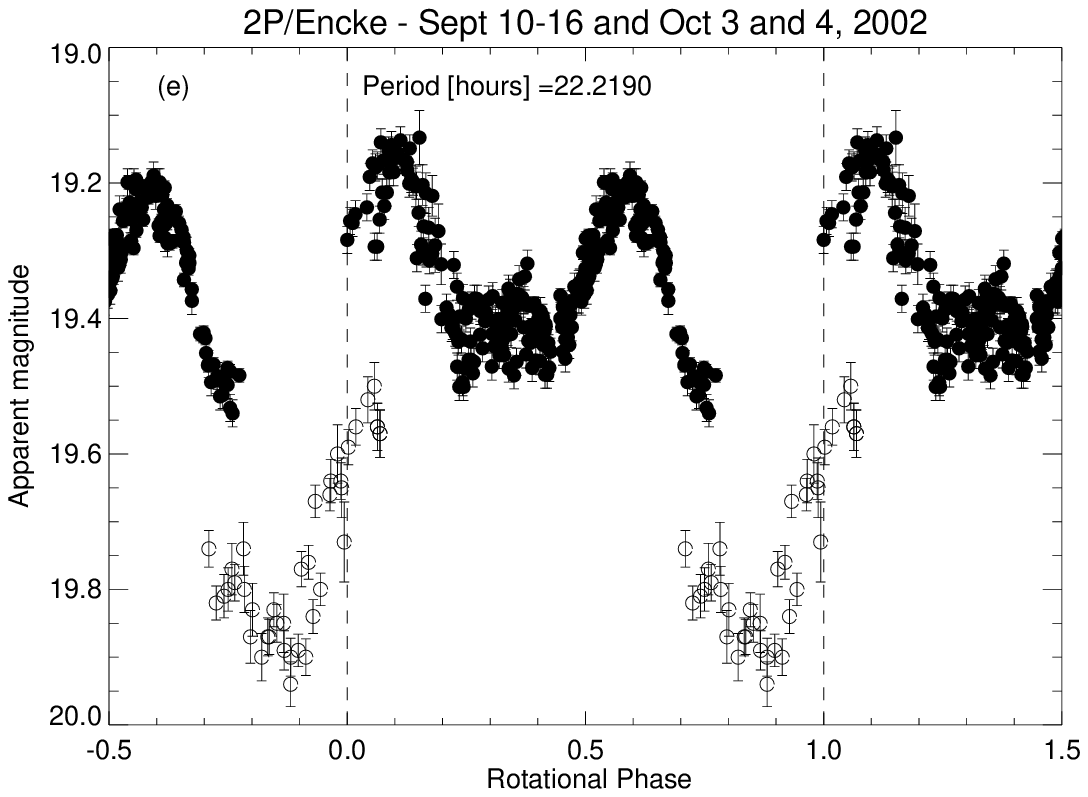}
}
\resizebox{(\hsize)/2}{!}
{
\includegraphics{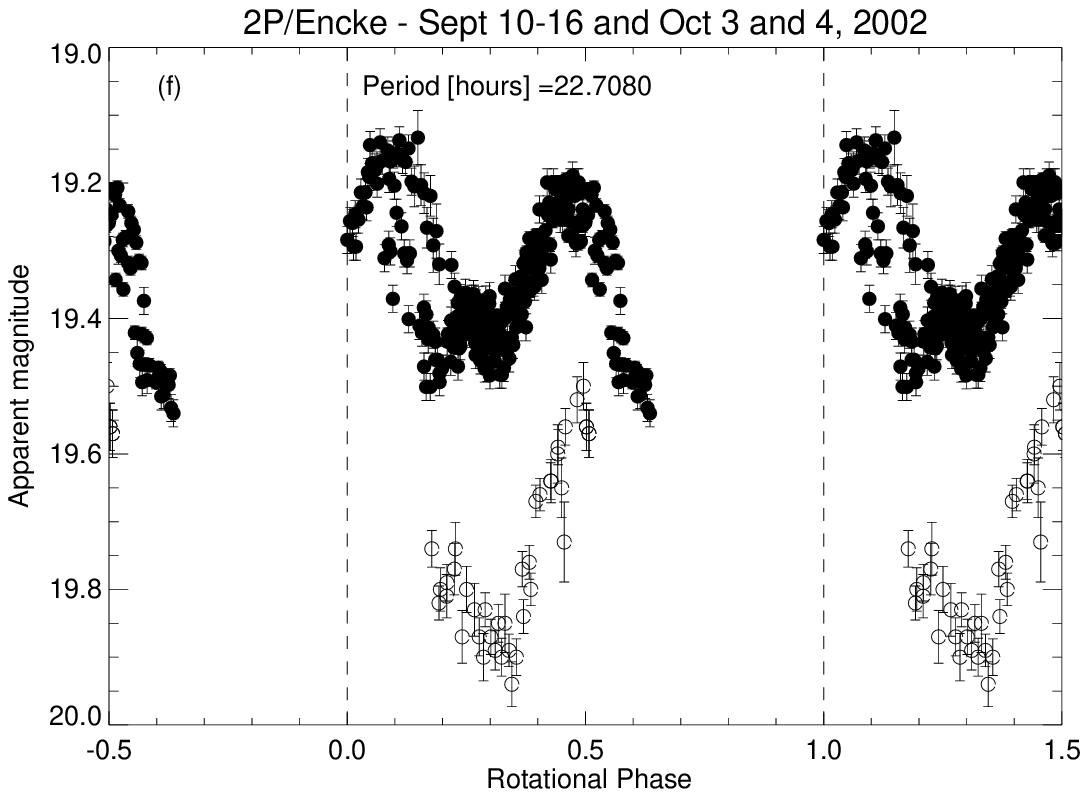}
}
\caption[]
{Our data from October 2002 was combined with $R$-band data from Fern\'{a}ndez et
al. (2005) spanning September 10-16, 2002, to improve the accuracy of the 
rotation period determination. (a) Several prominent features are seen in the periodogram,
with the most prominent and well defined being the period at $11.083 \pm 0.003$
hours. (b-f) The October and September data folded to each
of these five periodicities. The September data are shown as filled circles and
the October data are shown as open circles. 
}
\label{fig_YANsep_oct_phase_plots}
\end{figure}

The resultant periodogram from 2nd order fits is shown in 
Figure \ref{fig_YANsep_oct_phase_plots} (a). There are prominent, well defined 
features at 7.689, 11.083, and 15.153 hours, with a double feature at 
22.219 and 22.708 hours. As expected, there is no hint of a period near 16.30 hours 
(see section \ref{rotation_rate}). Plots (b-f) are the September and October data sets folded 
to each of the prominent periodicities.
We have left the magnitudes as apparent magnitudes to separate the two data
sets for clarity and to highlight the change in mean apparent magnitude between
the two dates. This brightness change is consistent with an inactive nucleus
on both occassions, but a tightly bound coma cannot be ruled out 
(see section \ref{Size_and_shape}).

One can see from Figure \ref{fig_YANsep_oct_phase_plots} (b) that 
a rotation period of 7.689 hours phases the October data set reasonably
well but does not work for the September data. In both cases 
the result is a physically unrealistic single-peaked lightcurve.
The phased data for the 15.153 hour period is somewhat interesting in that
although the folded data for September is quite poor, there is a hint of
similarity in lightcurve shape between the two data sets (i.e. one peak followed
by a slightly larger peak).
A rotation period of $11.089 \pm 0.003$ hours is clearly the most
likely period from the periodogram in 
Figure \ref{fig_YANsep_oct_phase_plots} (a). The scatter in the September data
is reduced significantly and the October brightness maxima and minima line up
extremely well.
This period agrees very well with the Fern\'{a}ndez et al. value of
$11.079 \pm 0.009$, although now the synodic period is known to a 
slightly higher degree of accuracy.

This raises an interesting point about our period determination
in section \ref{rotation_rate} of $11.45 \pm 0.11$ hours, as well as 
determining rotation periods in general from data taken over a few nights.
One can see that the 11.45~hour value is somewhat larger that the 
refined value of $11.089 \pm 0.003$ hours. Although
the two values are still consistent at the 3.3$\sigma$ level,
we feel that the difference is still sufficiently large to make the point 
that one must proceed with caution when extracting rotation information from
lightcurves based on only 2--3 nights, and time-bases on the order of weeks may be 
needed to get a solid synodic period value. In this case the true synodic period 
was at the extreme of the uncertainty range of our 2-day data set.

Fern\'{a}ndez et al. discussed the possibility of a period around 22.158
hours but that period doesn't show up in the periodogram of the combined data set.
The nearest feature to this period is the double feature at 
22.219 and 22.708 hours.
This double feature appears as a result of the reduced rotational-phase
coverage of the October data (from just two nights data as opposed to seven nights
in the September data set). Almost-equally good fits can be achieved by simply
shifting the phased October data left or right to line up with either of the two
peaks in the September phase plot. This is more clearly seen in figures 
\ref{fig_YANsep_oct_phase_plots}(e) and (f).
The September phasing with P=22.708 hours is not good, but is slightly
better for P=22.219 hours. In neither case do the peaks line up as well
as for the 11.089 hour value nor are the fit residuals as low. 
We therefore conclude that the
best solution for the combined data set is $11.089 \pm 0.003$ hours.

%%%%%%%%%%%%%%%%%%%%%%%%%%%%%%%%%%%%%%%%%%%%%%%%%%%%%%%%%%%%%%%%%%%%%%%%

\subsection{\bf Comparison with published rotation periods}
\label{Comparison_with_pub}

\begin{figure}[t]
\resizebox{(\hsize)/2}{!}
{
\includegraphics{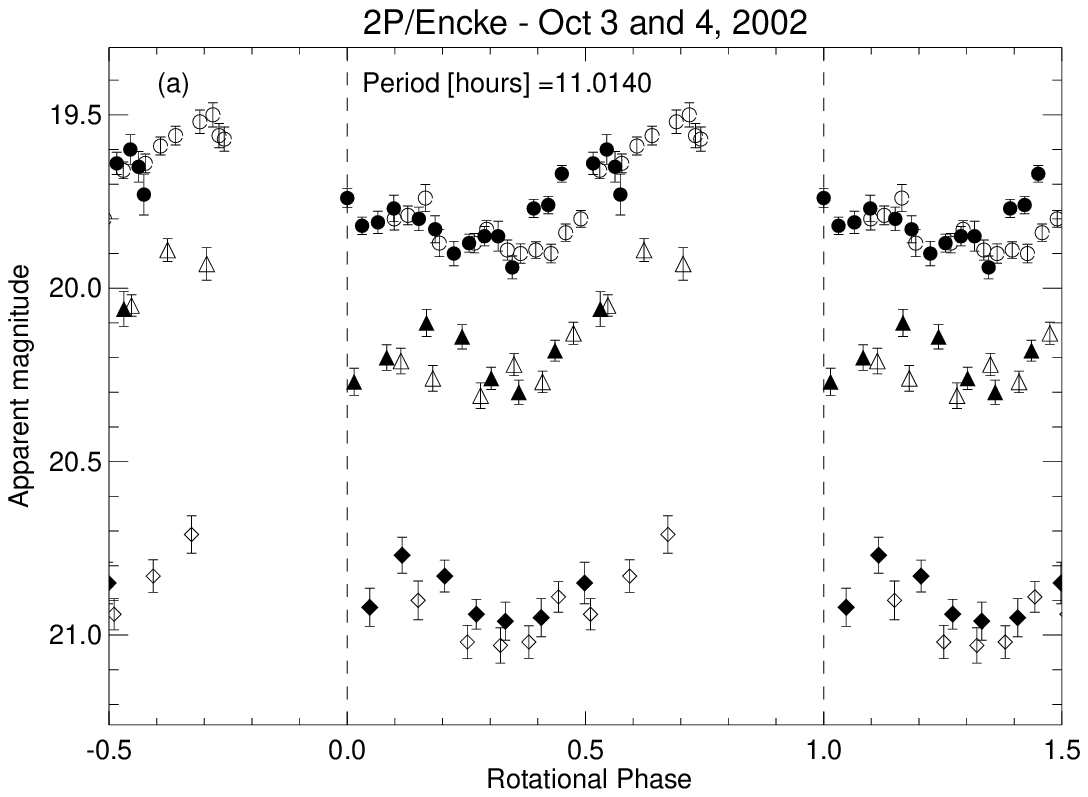}
}
\resizebox{(\hsize)/2}{!}
{
\includegraphics{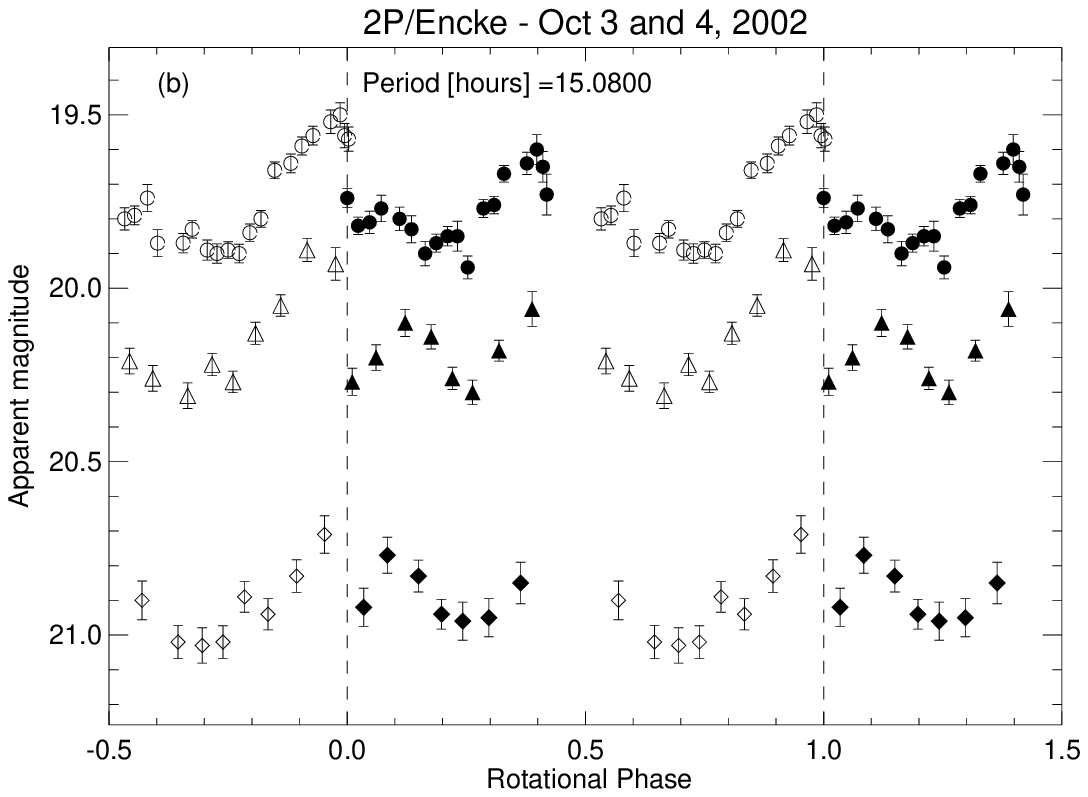}
}
\resizebox{(\hsize)/2}{!}
{
\includegraphics{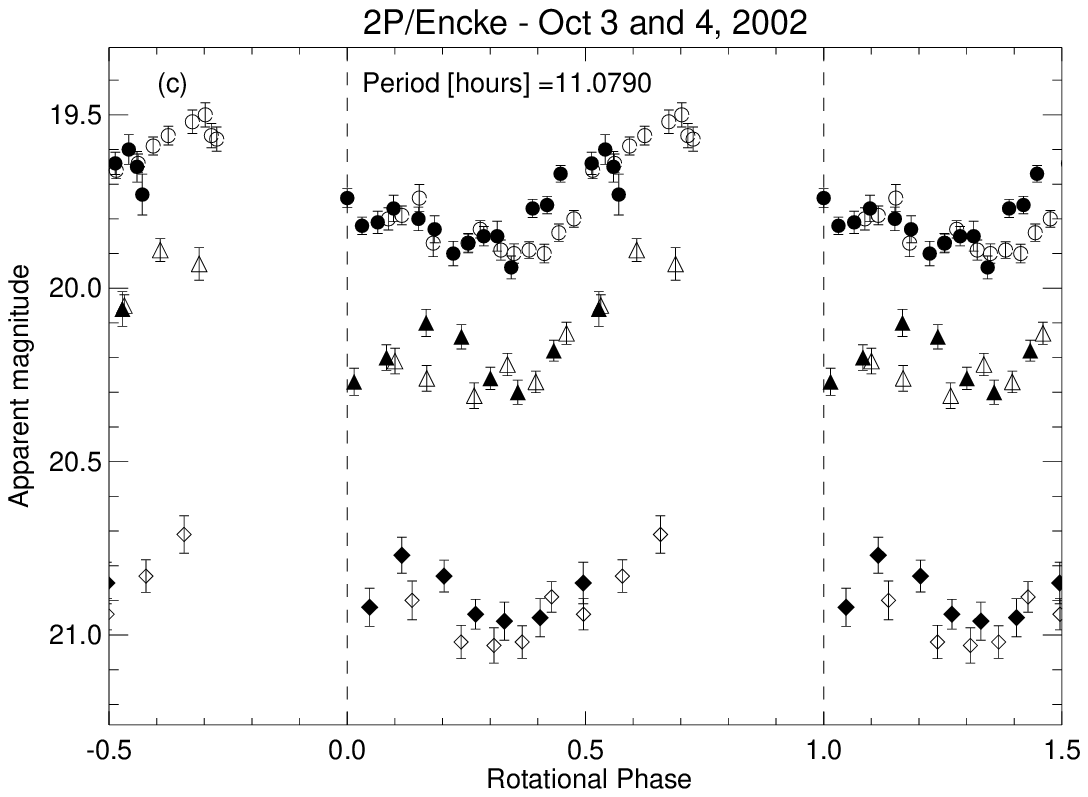}
}
\resizebox{(\hsize)/2}{!}
{
\includegraphics{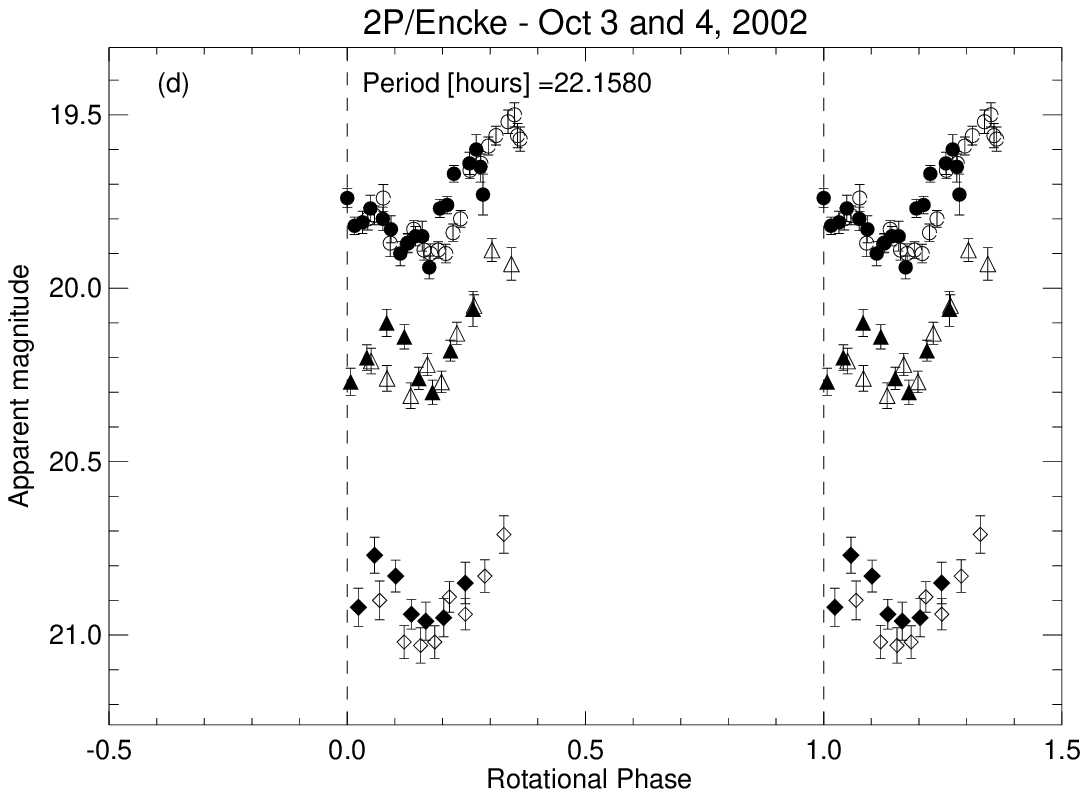}
}
\caption[]
{Comparison of our photometry with published periods.
(a) Lowry et al. (2003) - subset of Fernandez et al. (2005),
(b) Fernandez et al. (2000), Luu and Jewitt (1990),
(c) and (d) Fernandez et al. (2005). Symbol notation same as Figure \ref{fig_oct_phase_plots}.
}
\label{fig_published_phase_plots}
\end{figure}

As noted in previous sections, there are multiple lightcurves that have been
published, including the large data set of Fern\'{a}ndez et al. (2005). 
In section \ref{rotation_rate}, where we consider just our October 2002 data, 
we have arguments that support a rotation period at $11.45 \pm 0.11$ hours. This period
produces the best possible fit to the $R$-filter data but results in an odd-looking 
lightcurve that is difficult to explain in terms of shape, and does not 
work well if the lightcurves from all three filters are considered
collectively. This period disagrees with the Luu and Jewitt (1990) 
best-fit period of $15.08 \pm 0.08$ hours and the thermal infrared 
observations by Fern\'{a}ndez et al. (2000) which are consistent with the Luu and Jewitt 
period. Jewitt and Meech (1987) quote a most likely period of $22.43 \pm 0.08$ hours.
However, our result agrees well with the more recent Fern\'{a}ndez et al. (2005) data sets
which place the spin rate at either $11.079 \pm 0.009$ or
$22.158 \pm 0.012$ hours. In Figure \ref{fig_published_phase_plots} we phase our 
October data to each of the previous periods, including the 11.014 hour period from 
Lowry et al. (2003), which is a subset of the Fern\'{a}ndez et al. (2005) data.

Despite the obvious differences in the best-fit periods across the various
data sets, there are glaring commonalities in that $all$ the periodograms show
prominent features in the ranges $\sim$7-8, $~\sim$11-12, $\sim$15-16 and $\sim$22-23 hours.
The actual rotation rate must fall in one of these ranges, and we conclude that
the period just over 11 hours in the correct one, at least for the most recent data sets,
and assuming simple relaxed principle-axis rotation. Not only is this the best-fit 
period for the September 2002 data set (Fern\'{a}ndez et al. 2005) and our October 2002 data
set, but when they are linked together, it clearly offers the best solution. The period 
at 15.153 hours for the combined data set (Figure \ref{fig_YANsep_oct_phase_plots}) does not
phase well at all. The only negative argument regarding the $\sim$11 hour
period is the persistent misalignment of the brightness peaks when the data are phased to this
period. An important point is that new radar data (Harmon and Nolan 2005) supports the 
$\sim$11 hour rotation period, and $excludes$ the longer 15 and 22 hour 
periods. Invoking complex rotation, as suggested by Belton et al. (2005) 
(see section \ref{INTRODUCTION}), in conjunction with a highly elongated 
nucleus with significant surface inhomogenieties, may well be the only way to solve 
the spin-state problem of Encke. However, without futher data that would enable the necessary
modelling to take place, we choose the single-axis-rotation approach for now 
and arrive at a best-fit
synodic period of $11.083 \pm 0.003$ hours.

%%%%%%%%%%%%%%%%%%%%%%%%%%%%%%%%%%%%%%%%%%%%5

%%%%%%%%%%%%%%%%%%%%%%%%%%%%%%%%%%%%%%%%%%%%5
\noindent
\section{\bf Summary and main conclusions}
  \label{SUMMARY}

We present results from simultaneous $B$, $V$ and $R$-filter photometry obtained on October 3-4, 2002 at 
the Steward Observatory 2.3m Bok telescope on Kitt Peak. Rotational lightcurves in
all three filters were extracted and analysed to study the physical and color
properties of the nucleus. The main conclusions from this work are as follows.

\begin{enumerate}

\item The average color indices were measured for both nights and the values are
very similar. The values are
($V-R$) = $0.40 \pm 0.06$ and ($B-V$) = $0.70 \pm 0.08$ for October 3, 2002 
($\bar{R}$ = $19.78 \pm 0.03$);
and 
($V-R$) = $0.38 \pm 0.05$ and ($B-V$) = $0.76 \pm 0.03$  for October 4, 2002
($\bar{R}$ = $19.74 \pm 0.03$). Encke's colors are 
slightly redder than the equivalent Solar values, but are bluer than most other
comets as well as previous values reported for Encke.

\item On October 3 we see a clear and systematic dip in ($V-R$) color index, 
which is not seen for ($B-V$) on the same night, and is not repeated on October 4.
This systematic variation is considered real at the 2$\sigma$ confidence level.
However, for a nucleus rotation period of 11.089 hours, color variations
should be repeated from night to night. To explain the color variation, we explore the
possibility that gaseous emissions from C$_2$ molecules at $B$ and $V$ wavelengths 
occurred around Oct. 3.2 UT, but this is ruled out based on gas-expansion-velocity 
arguments.
Also, no coma or tail was seen in our deep images 
of the comet, nor in previous near aphelion imaging. Thus, we tend to favor the
no-coma scenario.
Complex rotation, in conjunction with a highly elongated nucleus with significant surface 
inhomogenieties, may well explain both the color variations
and the erratic photometric behaviour of the comet when it is near aphelion.

\item The average apparent $R$-band magnitude across $both$ nights is
$19.76 \pm 0.03$, which corresponds to a mean effective radius of
$3.95 \pm 0.06$ km. This value is similar to that found for the $V$- and $B$-filter 
photometry. Taking the observed brightness range, we get 
$a/b \geq 1.44 \pm 0.06$ for the axial ratio of Encke's nucleus. 
Applying this axial ratio to the $R$-filter photometry gives nucleus semi-axes of 
[$3.60 \pm 0.09$]$\times$[$5.20 \pm 0.13$] km, for an inactive nucleus. 
These size measurements use the empirically derived values for the geometric
albedo and phase coefficient of 0.047 and 0.06 magnitudes/degree, respectively
(Fern\'{a}ndez et al. 2000). 

\item We analysed the $R$-filter time-series photometry using the method
of Harris et al. (1989) to constrain the rotation period of the comet's nucleus.
We find that a period of $11.45 \pm 0.11$ hours satisfies the data, however the errors bars
are large.
A rotation period somewhere between 15-16 hours can work 
for the current data set in terms of the resulting lightcurve shape, which is reasonably 
consistent across the three observed wavelengths, and would explain the color-variation 
difference between each night. However, we believe that solutions in the 15-16 hour range 
are pathologic cases that brings the data from both nights together with no overlap.

\item It is clear that a period just over 11 hours is the most
likely value as this consistently produces by far the best fits 
(cf. Fern\'{a}ndez et al. 2005: Belton et al. 2005), and is supported by recent radar
observations (Harmon and Nolan 2005).
We have successfully linked our data with the September 2002 data set from 
Fernandez et al. (2005) - taken just 2-3 weeks
before the current data set - and we show that a rotation period of just over 11 hours does
indeed work very well for the combined data set. The resulting best-fit period is
$11.083 \pm 0.003$, consistent with the Fern\'{a}ndez et al. (2005) value.

\end{enumerate}

%%%%%%%%%%%%%%%%%%%%%%%%%%%%%%%%%%%%%%%%%%%%5
\noindent
{\bf Acknowledgments}

We thank H.~Campins and another anonymous referee for their very helpful reviews.
This work was performed in part while the first author held a National Research 
Council Associateship Award, and in part at Queen's University Belfast. Also, this work 
was performed in part at the Jet Propulsion 
Laboratory under contract with NASA. We acknowledge additional support from the NASA 
Planetary Astronomy Program. We thank Steward Observatory for granting time on the 
2.3m Bok Telescope.
IRAF is distributed by the National Optical Astronomy Observatories, which 
is operated by the Association of Universities for Research in Astronomy, 
Inc. (AURA) under cooperative agreement with the National Science Foundation.
We acknowledge JPL's Horizons online ephemeris generator for providing the
comet's position and rate of motion during the observations.\\
%%%%%%%%%%%%%%%%%%%%%%%%%%%%%%%%%%%%%%%%%

\noindent
{\bf References}

\def\paper#1#2#3#4#5#6{\hangindent=3pc \hangafter=1#1. #2. #3 {\it #4} {\bf #5}, #6.}

\def\paperinprep#1#2#3#4#5{\hangindent=3pc \hangafter=1#1. #2. #3 {\it #4.} #5.}

\def\paperinprepB#1#2#3#4{\hangindent=3pc \hangafter=1#1 #2. In {\it#3.} #4.}

\def\thesis#1#2#3#4{\hangindent=3pc \hangafter=1#1 #2. #3 #4.}

\def\book#1#2#3#4#5#6#7{\hangindent=3pc \hangafter=1#1. #2. #3. In {\it #4} (#5), #6. pp. #7.}

\def\booksingleauthor#1#2#3#4#5{\hangindent=3pc \hangafter=1#1 #2. In {\it #3}, pp. #4. #5.}

\def\proceedings#1#2#3#4#5#6#7{\hangindent=3pc \hangafter=1#1 #2. #3 In {\it #4} #5 {\bf #6}, #7.}

\def\proceedingsA#1#2#3#4#5#6{\hangindent=3pc \hangafter=1#1 #2. #3 In {\it #4}, #5, #6.}

\def\irafmanuals#1#2#3#4{\hangindent=3pc \hangafter=1#1 #2. #3. #4.}

\def\iauc#1#2#3#4{\hangindent=3pc \hangafter=1#1 #2. #3. #4.}

\def\mpec#1#2#3#4{\hangindent=3pc \hangafter=1#1. #2. {\it #3} {#4}.}

\noindent
\paper{A'Hearn, M.F., Belton, M.J.S., Delamere, W.A., Kissel, J., 
Klaasen, K.P., McFadden, L.A., Meech, K.J., Melosh, H.J., Schultz, P.H., 
Sunshine, J.M., and 23 co-authors}{2005}
{Deep Impact: Excavating Comet Tempel 1.}
{Science}{310}{258--264}

\noindent
\paper{Belton, M.J.S., Samarasinha, N.H., Fern\'{a}ndez, Y.R. and Meech, K.J}{2005}
{The excited spin state of Comet 2P/Encke.}
{Icarus}{175}{181--193} 

\noindent
\paper{Buratti, B.J., Hicks, M.D., Soderblom, L.A., Britt, D., Oberst, J. and Hillier, J.K}{2004}
{Deep Space 1 photometry of the nucleus of Comet 19P/Borrelly.}
{Icarus}{167}{16--29} 

\noindent
\paper{Duncan, M.J., Levison, H.F. and Budd, S.M}{1995}
{The Dynamical Structure of the Kuiper Belt.}
{AJ}{110}{3073--3081}

\noindent
\paper{Fern\'{a}ndez, Y.R., Lisse, C.M., Ulrich Kaufl, H., Peschke, 
S.B., Weaver, H.A., A'Hearn, M.F., Lamy, P.P., 
Livengood, T.A. and Kostiuk, T}{2000}
{Physical Properties of the Nucleus of Comet 2P/Encke.}
{Icarus}{147}{145--160}

\noindent
\paper{Fernandez, Y.R., Lowry, S.C., Weissman, P.R., Mueller, B.E.A., 
Samarasinha, N.H., Belton, M.J.S. and Meech, K.J}{2005}
{New near-aphelion light curves of comet 2P/Encke.}
{Icarus}{175}{194--214}

\noindent
\paper{Harmon, J.K. and Nolan, M.C}{2005}
{Radar observations of Comet 2P/Encke during the 2003 apparition.}
{Icarus}{176}{175--183}

\noindent
\paper{Harris, A.W., Young, J.W., Bowell, E., Martin, L.J., Millis, R.L.,
Poutanen, M., Scaltriti, F., Zappala, V., Schober, H.J., 
Debehogne, H., and Zeigler, K.W}
{1989}{Photoelectric observations of asteroids 3, 24, 60, 261, and 863.}
{Icarus}{77}{171--186}

\noindent
\paper{Ip, W.H. and Fern\'{a}ndez, J.A}{1997}
{On dynamical scattering of Kuiper Belt Objects in 2:3 resonance with Neptune 
into short-period comets.}
{A\&A}{324}{778--784} 

\noindent
\paper{Jewitt, D. and Meech, K}{1987}
{CCD photometry of Comet P/Encke.}
{AJ}{93}{1542--1548}	

\noindent
\paper{Kamoun, P.G., Campbell, D.B., Ostro, S.J., Pettengill, G.H. and Shapiro, I.I}{1982}
{Comet Encke - Radar detection of nucleus.}
{Science}{216}{293--295}

\noindent
\book{Lamy, P.L., Toth, I., Fern\'{a}ndez Y.R. and Weaver H.A}{2004} 
{The sizes, shapes, albedos, and colors of cometary nuclei}
{Comets II}
{M. Festou, H.U. Keller and H.A. Weaver, Editors} 
{Univ. of Arizona Press, Tucson}{223--264}

\noindent
\paper{Landolt, A.U}{1992}
{UBVRI photometric standard stars in the magnitude range 11.5-16.0 around the celestial equator.}
{AJ}{104}{340--371}

\noindent
\paper{Lowry, S.C. and Fitzsimmons, A}{2005}
{WHT observations of distant comets.}
{MNRAS}{358}{641--650}

\noindent
\paper{Lowry, S.C., Weissman, P.R., Sykes, M.V. and Reach, W.T}{2003}
{Observations of periodic comet 2P/Encke: Physical properties of 
the nucleus and first visual-wavelength detection of its dust trail.}
{LPSC}{34}{2056}

\noindent
\paper{Luu, J. and Jewitt, D}{1990}
{The nucleus of Comet P/Encke}	
{Icarus}{86}{69--81} 

\noindent
\book{Samarasinha, N.H., Mueller, B.E.A, Belton, M.J.S. and Jorda, L}{2004} 
{Rotation of cometary nuclei}
{Comets II}
{M. Festou, H.U. Keller and H.A. Weaver, Editors} 
{Univ. of Arizona Press, Tucson}{281--300}

\noindent
\proceedingsA{Sarugaku, Y., Ishiguro, M., Miura, N., Usui, F., and Ueno, M.}{2005}
{Optical Observations of the Comet 2P/Encke Dust Trail.}
{Proceedings of Dust in Planetary Systems}{LPI Contribution No. 1280}{p127}

\noindent
\proceedings{Tody, D.}{1986}
{The IRAF data reduction and analysis system.}
{Proc. SPIE Instrumentation in Astronomy VI}
{(D.L. Crawford, Eds.)}{627}{733}

\noindent
\proceedings{Tody, D.}{1993}
{IRAF in the Nineties.}
{Astronomical Data Analysis Software and Systems II, A.S.P. Conference Series}
{(R.J. Hanisch, R.J.V. Brissenden, and J. Barnes, Eds.)}{52}{173}

\noindent
\book{Weissman, P.R., Asphaug, E. and Lowry, S.C}{2004} 
{Structure and density of cometary nuclei}
{Comets II}
{M. Festou, H.U. Keller and H.A. Weaver, Editors} 
{Univ. of Arizona Press, Tucson}{337--358}

\clearpage

% TABLES  %%%%%%%%%%%%%%%%%%

%TABLE 1
\begin{table*}
\caption[]
{\textbf{Geometry of comet during observations}}
\label{Obs_geom_table}
\begin{center}
\begin{tabular}{cccc}
\hline
\\
UT Date &$R_{h}$ [AU] &$\Delta$ [AU] &$\alpha$ [deg.]     
\\
\\				     
\hline
\\
Oct 3, 2002  & 3.925--3.925  & 3.020--3.021  & 7.09--7.16   \\												  
Oct 4, 2002  & 3.923--3.922  & 3.025--3.027  & 7.35--7.43   \\
\\												  
\hline  			        							  
\end{tabular}											  
\end{center}											  
\end{table*}

%TABLE 2
\begin{table}[t]
\caption[]
{\textbf{Observation log and $B$, $V$ and $R$-filter photometry}}
\label{Relative_mags_table}
\begin{center}
\begin{tabular}{cccccc}
\hline
\\
UT$^{\dagger}$ &$m_{\lambda}\pm\sigma$  &Filter &t     &($V-R$) $\pm$ $\sigma$ 	&($B-V$) $\pm$ $\sigma$
\\
(midtime)      &			&       &[sec] &			&	 
\\
\\							     
\hline
\\
\multicolumn{6}{l}{\textbf{October 3, 2002}}\\

3.1183  &19.74 $\pm$ 0.03 &R &400 &0.53 $\pm$ 0.05 &		    \\
3.1249  &20.27 $\pm$ 0.04 &V &400 &0.49 $\pm$ 0.05 &0.65 $\pm$ 0.07 \\
3.1326  &19.82 $\pm$ 0.03 &R &400 &0.45 $\pm$ 0.05 &		    \\
3.1399  &20.92 $\pm$ 0.06 &B &500 &0.42 $\pm$ 0.07 &0.68 $\pm$ 0.08 \\
3.1478  &19.81 $\pm$ 0.03 &R &400 &0.40 $\pm$ 0.05 &		    \\
3.1563  &20.20 $\pm$ 0.04 &V &400 &0.41 $\pm$ 0.06 &0.64 $\pm$ 0.08 \\
3.1631  &19.77 $\pm$ 0.04 &R &400 &0.43 $\pm$ 0.05 &		    \\
3.1711  &20.77 $\pm$ 0.05 &B &500 &0.37 $\pm$ 0.07 &0.62 $\pm$ 0.07 \\
3.1872  &19.80 $\pm$ 0.03 &R &400 &0.30 $\pm$ 0.05 &		    \\
3.1946  &20.10 $\pm$ 0.04 &V &400 &0.29 $\pm$ 0.06 &0.70 $\pm$ 0.08 \\
3.2029  &19.83 $\pm$ 0.04 &R &400 &0.27 $\pm$ 0.06 &		    \\
3.2121  &20.83 $\pm$ 0.05 &B &500 &0.26 $\pm$ 0.07 &0.71 $\pm$ 0.07 \\
3.2210  &19.90 $\pm$ 0.04 &R &400 &0.24 $\pm$ 0.05 &		    \\
3.2289  &20.14 $\pm$ 0.04 &V &400 &0.25 $\pm$ 0.06 &0.75 $\pm$ 0.07 \\
3.2356  &19.87 $\pm$ 0.03 &R &400 &0.27 $\pm$ 0.04 &		    \\
3.2426  &20.94 $\pm$ 0.04 &B &500 &0.34 $\pm$ 0.06 &0.74 $\pm$ 0.06 \\
3.2505  &19.85 $\pm$ 0.03 &R &400 &0.41 $\pm$ 0.04 &		    \\
3.2569  &20.26 $\pm$ 0.03 &V &400 &0.41 $\pm$ 0.06 &0.69 $\pm$ 0.08 \\
3.2636  &19.85 $\pm$ 0.04 &R &400 &0.41 $\pm$ 0.05 &		    \\
3.2706  &20.96 $\pm$ 0.06 &B &500 &0.38 $\pm$ 0.07 &0.68 $\pm$ 0.07 \\
3.2772  &19.94 $\pm$ 0.03 &R &400 &0.36 $\pm$ 0.05 &		    \\
3.2835  &20.30 $\pm$ 0.04 &V &400 &0.44 $\pm$ 0.05 &0.66 $\pm$ 0.09 \\
3.2979  &19.77 $\pm$ 0.03 &R &400 &0.52 $\pm$ 0.04 &		    \\
3.3051  &20.95 $\pm$ 0.06 &B &500 &0.47 $\pm$ 0.06 &0.71 $\pm$ 0.07 \\
3.3119  &19.76 $\pm$ 0.03 &R &400 &0.42 $\pm$ 0.04 &		    \\
3.3184  &20.18 $\pm$ 0.03 &V &400 &0.47 $\pm$ 0.05 &0.72 $\pm$ 0.09 \\
3.3250  &19.67 $\pm$ 0.02 &R &400 &0.52 $\pm$ 0.04 &		    \\
3.3469  &20.85 $\pm$ 0.06 &B &500 &0.47 $\pm$ 0.07 &0.73 $\pm$ 0.08 \\
3.3551  &19.64 $\pm$ 0.03 &R &400 &0.42 $\pm$ 0.06 &		    \\
3.3620  &20.06 $\pm$ 0.05 &V &400 &0.44 $\pm$ 0.07 &0.79 $\pm$ 0.08 \\
3.3682  &19.60 $\pm$ 0.04 &R &400 &0.46 $\pm$ 0.07 &		    \\
3.3762  &19.65 $\pm$ 0.04 &R &400 &		   &		    \\
3.3813  &19.73 $\pm$ 0.06 &R &400 &		   &		    \\

\\												  
\hline  			        							  
\end{tabular}											  
\end{center}											  
\end{table}

\clearpage
\addtocounter{table}{-1}

\begin{table}[t]
\caption[]
{\textbf{\textit{Continued:} Observation log and $BVR$ photometry}}
\begin{center}
\begin{tabular}{cccccc}
\hline
\\
UT$^{\dagger}$ &$m_{\lambda}\pm\sigma$  &Filter &t     &($V-R$) $\pm$ $\sigma$ 	&($B-V$) $\pm$ $\sigma$
\\
(midtime)      &			&       &[sec] &			&	 
\\
\\							     
\hline
\\
\multicolumn{6}{l}{\textbf{October 4, 2002}}\\

4.0814  &19.80 $\pm$ 0.03 &R &400 &0.41 $\pm$ 0.05 &	            \\
4.0878  &20.21 $\pm$ 0.04 &V &400 &0.41 $\pm$ 0.06 &0.69 $\pm$ 0.07 \\
4.0944  &19.80 $\pm$ 0.03 &R &400 &0.42 $\pm$ 0.05 &	            \\
4.1043  &20.90 $\pm$ 0.06 &B &500 &0.47 $\pm$ 0.07 &0.66 $\pm$ 0.08 \\
4.1113  &19.74 $\pm$ 0.04 &R &400 &0.52 $\pm$ 0.05 &	            \\
4.1185  &20.26 $\pm$ 0.04 &V &400 &0.46 $\pm$ 0.07 &0.70 $\pm$ 0.08 \\
4.1248  &19.87 $\pm$ 0.04 &R &400 &0.40 $\pm$ 0.05 &	            \\
4.1518  &21.02 $\pm$ 0.05 &B &500 &0.41 $\pm$ 0.07 &0.73 $\pm$ 0.07 \\
4.1586  &19.87 $\pm$ 0.03 &R &400 &0.43 $\pm$ 0.05 &	            \\
4.1645  &20.31 $\pm$ 0.04 &V &400 &0.45 $\pm$ 0.05 &0.72 $\pm$ 0.08 \\
4.1704  &19.83 $\pm$ 0.03 &R &400 &0.47 $\pm$ 0.04 &	            \\
4.1837  &21.03 $\pm$ 0.05 &B &500 &0.40 $\pm$ 0.06 &0.77 $\pm$ 0.07 \\
4.1903  &19.89 $\pm$ 0.03 &R &400 &0.32 $\pm$ 0.04 &	            \\
4.1967  &20.22 $\pm$ 0.03 &V &400 &0.32 $\pm$ 0.05 &0.81 $\pm$ 0.08 \\
4.2031  &19.90 $\pm$ 0.03 &R &400 &0.32 $\pm$ 0.04 &	            \\
4.2110  &21.02 $\pm$ 0.05 &B &500 &0.35 $\pm$ 0.06 &0.78 $\pm$ 0.06 \\
4.2177  &19.89 $\pm$ 0.02 &R &400 &0.37 $\pm$ 0.04 &	            \\
4.2242  &20.27 $\pm$ 0.03 &V &400 &0.37 $\pm$ 0.05 &0.69 $\pm$ 0.07 \\
4.2324  &19.90 $\pm$ 0.03 &R &400 &0.36 $\pm$ 0.04 &	            \\
4.2395  &20.89 $\pm$ 0.04 &B &500 &0.33 $\pm$ 0.06 &0.69 $\pm$ 0.06 \\
4.2465  &19.84 $\pm$ 0.03 &R &400 &0.30 $\pm$ 0.04 &	            \\
4.2539  &20.13 $\pm$ 0.03 &V &400 &0.31 $\pm$ 0.05 &0.78 $\pm$ 0.07 \\
4.2611  &19.81 $\pm$ 0.02 &R &400 &0.33 $\pm$ 0.04 &	            \\
4.2704  &20.94 $\pm$ 0.05 &B &500 &0.36 $\pm$ 0.06 &0.85 $\pm$ 0.06 \\
4.2793  &19.66 $\pm$ 0.02 &R &400 &0.39 $\pm$ 0.04 &	            \\
4.2872  &20.05 $\pm$ 0.03 &V &400 &0.40 $\pm$ 0.05 &0.83 $\pm$ 0.07 \\
4.3005  &19.64 $\pm$ 0.03 &R &400 &0.41 $\pm$ 0.04 &	            \\
4.3081  &20.83 $\pm$ 0.05 &B &500 &0.36 $\pm$ 0.06 &0.86 $\pm$ 0.07 \\
4.3151  &19.59 $\pm$ 0.03 &R &400 &0.31 $\pm$ 0.04 &	            \\
4.3219  &19.89 $\pm$ 0.03 &V &400 &0.32 $\pm$ 0.05 &0.88 $\pm$ 0.08 \\
4.3297  &19.56 $\pm$ 0.03 &R &400 &0.34 $\pm$ 0.04 &	            \\
4.3450  &20.71 $\pm$ 0.05 &B &400 &0.37 $\pm$ 0.07 &0.80 $\pm$ 0.08 \\
4.3531  &19.52 $\pm$ 0.03 &R &400 &0.41 $\pm$ 0.06 &	            \\
4.3596  &19.93 $\pm$ 0.05 &V &500 &0.42 $\pm$ 0.07 &0.78 $\pm$ 0.07 \\
4.3656  &19.50 $\pm$ 0.04 &R &500 &0.43 $\pm$ 0.06 &	            \\
4.3715  &19.56 $\pm$ 0.04 &R &400 &	           &	            \\
4.3766  &19.58 $\pm$ 0.04 &R &400 &	           &	            \\
\\												  
\hline  			        							  
\end{tabular}											  
\end{center}											  
\begin{small}
$\dagger$  Light time corrected\\
The relative magnitudes have been calibrated and placed on a standard 
magnitude scale (Landolt 1992).
\end{small}  
\end{table}

%TABLE 3
\begin{table*}
\caption[]
{\textbf{Mean effective radii from $BVR$ photometry}}
\label{Sizes_table}
\begin{center}
\begin{tabular}{cccc}
\hline
\\
UT Date &Filter &$m_{\lambda}\pm\sigma$  &Radius $\pm\sigma$ [km]     
\\
\\				     
\hline
\\
Oct 3, 2002   &R   &19.78 $\pm$ 0.03  &3.89 $\pm$ 0.06 \\											   
              &V   &20.19 $\pm$ 0.04  &3.80 $\pm$ 0.07 \\											   
              &B   &20.89 $\pm$ 0.05  &3.74 $\pm$ 0.09 \\
\\			        	   	        						   
Oct 4, 2002   &R   &19.74 $\pm$ 0.03  &4.01 $\pm$ 0.05 \\											   
              &V   &20.14 $\pm$ 0.04  &3.93 $\pm$ 0.06 \\											   
              &B   &20.92 $\pm$ 0.05  &3.74 $\pm$ 0.08 \\
\\												 
\hline  			        							  
\end{tabular}											  
\end{center}											  
\end{table*}

\clearpage

% FIGURES %%%%%%%%%%%%%%%%%%

%FIGURE 1

%FIGURE 2

%FIGURE 3

%FIGURE 4

%FIGURE 5

%FIGURE 6

%FIGURE 7

\end{document}